\documentclass[reprint,aip,jcp,superscriptaddress,floatfix]{revtex4-1}

\usepackage{amsmath,amssymb,txfonts}
\usepackage{graphicx}
\usepackage{color}

\allowdisplaybreaks

\newcommand{\lu}{\lambda_\text{u}}
\newcommand{\ls}{\lambda_\text{s}}
\newcommand{\omb}{\omega_\text{b}}
\newcommand{\omy}{\omega_y}
\newcommand{\kT}{k_\text{B}T}
\newcommand{\avg}[1]{\left\langle #1 \right\rangle}

\renewcommand{\Re}{\operatorname{Re}}
\renewcommand{\vec}[1]{\boldsymbol{#1}}

\begin{document}
\title{Reaction rate calculation with time-dependent invariant manifolds}
\author{Thomas Bartsch}
\affiliation{Department of Mathematical Sciences, Loughborough University,
Loughborough LE11 3TU, United Kingdom}
\author{F. Revuelta}
\affiliation{Grupo de Sistemas Complejos and Departamento de F\'isica y Mec\'anica,
Escuela T\'ecnica Superior de Ingenieros Agr\'onomos,
Universidad Polit\'ecnica de Madrid,
Avda. Complutense s/n 28040 Madrid, Spain}
\author{R. M. Benito}
\affiliation{Grupo de Sistemas Complejos and Departamento de F\'isica y Mec\'anica,
Escuela T\'ecnica Superior de Ingenieros Agr\'onomos,
Universidad Polit\'ecnica de Madrid,
Avda. Complutense s/n 28040 Madrid, Spain}
\author{F. Borondo}
\affiliation{Departamento de Qu\'imica, and
Instituto Mixto de Ciencias Matem\'aticas CSIC-UAM-UC3M-UCM,
Universidad Aut\'onoma de Madrid, Cantoblanco, 28049  Madrid, Spain}

\date{\today}

\begin{abstract}

The identification of trajectories that contribute to the reaction rate is the
crucial dynamical ingredient in any classical chemical reactivity calculation.
This problem often requires a full scale numerical simulation of the dynamics,
in particular if the reactive system is exposed to the influence of a heat bath.
As an efficient alternative, we propose here to compute invariant surfaces in
the phase space of the reactive system that separate reactive from nonreactive
trajectories.
The location of these invariant manifolds depends both on time and on the
realization of the driving force exerted by the bath.
These manifolds allow the identification of reactive trajectories
simply from their initial conditions, without the need of any further
simulation.
In this paper, we show how these invariant manifolds can be calculated,
and used in a formally exact reaction rate calculation based on perturbation
theory for any multidimensional potential coupled to a noisy environment.
\end{abstract}

\pacs{82.20.Db, 05.40.Ca, 05.45.2a, 34.10.+x}

\maketitle

\section{Introduction}

Transition State Theory (TST) provides the conceptual framework for large parts
of reaction rate theory.
Originally developed to describe the reactivity of small
molecules,\cite{Truhlar83,Truhlar96,Miller98}
it was later extended to encompass a
wide variety of processes in different branches of science, whose only commonality
is a transition from well-defined ``reactant'' to ``product''
states.\cite{Toller85,Eckhardt95,Hernandez93,Hernandez94,Jaffe99,Jaffe00,Koon00,Jaffe02,Uzer02,Komatsuzaki02}
The reason for this success is that TST proposes a simple answer to the two central
problems of reaction dynamics: It identifies a reaction mechanism, and provides
at the same time a simple approximation to the reaction rate.

More specifically,
TST is based on the observation that the rate limiting step in many reactions is the
crossing of an energetic barrier.
The top of this barrier then forms a bottleneck
in the phase space of the reactive system.
A reaction can only take place if the barrier is crossed.
If a dividing surface (DS) between reactant and product regions of phase space is placed close
to the bottleneck, the reaction rate can be computed from the steady-state flux
through that surface.
A strictly recrossing free DS can be constructed in the phase space
of reactive systems with arbitrarily many degrees of freedom
\cite{Uzer02,Waalkens04a,Waalkens04c}.
The simplest approximation to the rate is then obtained under the assumption that reactive
classical trajectories cross the DS only once and never return.
This assumption is often appropriate for reactions in the gas phase if the DS
is adequately chosen, but even then many reactions strongly violate this assumption.
Moreover,
if the system is strongly coupled to an environment, for example a liquid solvent,
the no-recrossing assumption is usually impossible to enforce strictly,
and often any DS is crossed many times by a typical trajectory.
As a result, a TST rate calculation significantly
overestimates the reaction rate.
For this reason, the focus of TST has long been to construct a DS that
eliminates or at least minimizes recrossings
(see Ref.~\onlinecite{Garrett05a} for a review).

The recrossing problem can be solved if
the reactive trajectories that contribute to the rate can be identified reliably.
An obvious means to this end is the
numerical simulation of representative trajectories
under the influence of the environment.
However, such calculations are usually very time consuming.
The advantage of the TST approximation is its simplicity.
It identifies reactive trajectories
simply by noting that they cross the DS from the reactant to the product side.
This criterion, which fails if recrossing cannot be ruled out,
is easy to use because it only takes account of the instantaneous velocity
with which a trajectory crosses the DS.
Nevertheless, it raises the prospect of a criterion to identify reactive trajectories
simply from their initial conditions, without the need to study their time evolution.
In the present paper we will derive such a criterion and demonstrate how it can be
used in a rate calculation.

The Langevin equation has been widely used to model the interaction of a reactive system
with a surrounding heat bath.\cite{Haenggi90,Pechukas76,Chandler78}
Being a classical model, this description neglects quantum effects
such as barrier tunnelling, which can be important in the case of
light particles \cite{Bothma10}, and the interaction with excited
surfaces through conical intersections \cite{Polli10}.
In this setting, Kramers\cite{Kramers40} explicit derived expressions for the rate of
escape across a parabolic barrier that apply separately in the limits of weak and strong damping.
The generalized Langevin equation is equivalent to a Hamiltonian model in which the reactive
system is bilinearly coupled to a bath of harmonic oscillators.\cite{Zwanzig73}
This reformulation allowed extensions of Kramers' rate theory that apply to situations with
arbitrary friction\cite{Melnikov86,Pollak89} or that include corrections due to anharmonic
barriers.\cite{Pollak93a,Talkner93,Talkner94a}
In this respect, it has long been predicted that the rates of activated processes should
rise with the coupling to the solvent in the weak coupling regime.
However, its direct observation in particle-based models had been elusive because
the coupling typically places the processes
in the spatial-diffusion limited regime wherein rates decrease with increasing friction.
Recently, the Kramers turnover in the rate with microscopic friction has been observed in
molecular dynamics trajectories calculation of the LiNC$\leftrightharpoons$LiCN in a bath
of Ar atoms.\cite{Muller10}
This observation provided direct and unambiguous evidence for the energy-diffusion
regime in which rates increase with friction.
In the present work we will not consider any explicit Hamiltonian model for the heat bath;
its influence will instead be described by means of a Langevin equation.
This approach allows to work within the finite-dimensional phase space of
the reactive system alone,
rather than the infinite-dimensional phase space of the bath.
This is advantageous from a computational point of view
and also conceptually convenient because the phase space
is easier to visualize in low dimension.

The aim of this paper is to describe the geometric
phase space structures that allow to classify a trajectory as reactive
or nonreactive just by looking at its initial condition,
thus avoiding the need of carrying out a numerical simulation.
Because the fate of a trajectory with a given initial condition depends on
the external force to which it is exposed, any such criterion must
take account of the precise realization of that force.
A general framework to do that was proposed in a recent series of
papers,\cite{Bartsch05b,Bartsch05c,Bartsch06a,Bartsch08,Hernandez10}
including the identification of
reactive trajectories~\cite{Bartsch06a}
and the rate calculation~\cite{Bartsch08}.
It was there shown that the Langevin equation gives rise to a specific
trajectory called the Transition State (TS) trajectory that remains
in the vicinity of the energetic barrier for all times,
without ever descending into any of the potential wells.
This TS trajectory depends on the realization of the noise,
and takes over the role of the fixed saddle point in the conventional TST.
A crucial observation in Refs.~\onlinecite{Bartsch05b,Bartsch05c} for the case
of a harmonic barrier is that the dynamics described by the Langevin equation
become noiseless when expressed in a time-dependent coordinate system
for which the TS~trajectory is the moving origin.
In the system of
relative coordinates  it is easy to identify a TST DS  that is rigorously free from recrossing.
It gives rise to a DS in the original, space fixed coordinate system
that is still recrossing-free.
This DS is time-dependent since it is
attached to the TS~trajectory, and it moves through phase space with it.
Even more significantly, this construction yields surfaces in phase space that
separate reactive from nonreactive trajectories.
These surfaces are the stable and unstable manifolds of the TS~trajectory,
and they also depend on time and on the realization of the noise.
Once they are known, initial conditions on one side of the surface are immediately
classified as reactive, while those on the other side are nonreactive.
Thus, the existence of these
invariant manifolds solves the diagnostic problem of standard rate theory
that was explained above.
They were used in Ref.~\onlinecite{Bartsch08} to obtain a compact rate formula,
strictly valid only for harmonic barriers.
An \emph{ad hoc} application to systems with an anharmonic barrier
produced, however, promising results.\cite{Bartsch06a,Bartsch08}

In the present paper, we develop a rigorous generalization of the
time-dependent TST formalism applicable to anharmonic barriers
using perturbation theory.
We show that the invariant manifolds persist in anharmonic systems and,
more importantly, they retain the ability to distinguish between
reactive and nonreactive trajectories, thus determining the chemical
reactivity of the system.
Finally, a simple perturbative scheme that allows one to calculate the invariant
manifolds for a specific anharmonic potential barrier will be presented,
and it will be used to obtain an analytic expansion for the reaction rate.
In the first part of the paper, we restrict our study to the one-dimensional case.
In this situation, the finite barrier corrections that were obtained in
Refs.~\onlinecite{Pollak93a,Talkner93,Talkner94a} will be recovered.
We have already given a brief account of these results in Ref.~\onlinecite{Revuelta12}.
We will here supply the details of the calculation that could not be presented within
the confines of a Communication. We will then introduce the  modifications
to the theory that are necessary to accommodate multidimensional reactive systems.
The efficacy of our method is demonstrated by deriving the first and second order
corrections to the reaction rate in the two-dimensional model potential already
used in Refs.~\onlinecite{Bartsch06a,Bartsch08}.

A final point is worth commenting on in this Introduction.
Perturbative rate calculations on multidimensional anharmonic barriers
have also been recently reported
in Refs.~\onlinecite{Kawai09,Kawai09a,Kawai10,Kawai10a}.
As in the present work, these authors based their work
on the identification of the TS~trajectory for the harmonic limit in
Refs.~\onlinecite{Bartsch05b,Bartsch05c}.
Our work, however, goes beyond those previous results in two main respects.
First, and most importantly, it provides an explicit and detailed description of the invariant
geometric structures in phase space that govern the reaction dynamics,
rather than studying them implicitly through approximate invariants and their imprint
on an ensemble of trajectories.
Second, whereas the normal form procedure in Refs.~\onlinecite{Kawai09,Kawai09a,Kawai10,Kawai10a}
aims at constructing a coordinate system in which the dynamics in the neighborhood
of the barrier can be simplified in general terms,
we derive a version of the perturbation theory that is specifically directed at
calculating the invariant manifolds that are relevant to reaction rate theory.
This perturbative scheme can therefore be much simpler,
and permits the analytical computation of corrections to Kramers' transmission factor
for anharmonic potentials.
Indeed, the calculation of the invariant manifolds can be easily carried out by hand,
whereas a normal form transformation always requires computer assitance.
This ease of computation makes the invariant manifolds an attractive tool for practical
rate calculations.

The outline of the paper is as follows.
In Section~\ref{sec:ratesIntro} we present the basic definitions and results of
rate theory that will be used to develop our method.
Section~\ref{sec:tdmf} is devoted to
a qualitative description of the invariant manifolds that give structure to the dynamics
in the vicinity of an energy barrier,
and section~\ref{sec:perturb1D} presents a method for their calculation.
In section~\ref{sec:rate} a general expression for the reaction rate in the case of
an anharmonic barrier is derived.
A description of the statistical properties of the invariant manifolds that are required
to evaluate the rate formula, the perturbative and numerical results for
various one-dimensional potentials are also given.
Finally, in section~\ref{sec:2d} we discuss the modifications
to the foregoing developments that are required in multidimensional systems,
and we also present results for the reaction rate on an anharmonic two-dimensional barrier.

\section{Fundamentals of rate theory}
\label{sec:ratesIntro}

In this section we outline the fundamentals of reaction rate theory that will be used in the
rest of the paper.
The reader is referred to Refs.~\onlinecite{Haenggi90, Pechukas76, Chandler78} for more details.

We assume that the reactant and product regions in configuration space are separated by a DS
that is characterized by the value $x=x^\ddag$ of a generalized reaction coordinate $x$,
which we choose such that the product region is given by $x>x^\ddag$.
The reaction rate is then given by the flux-over-population expression
\begin{equation}
    \label{kDef}
    k= \frac{J}{N},
\end{equation}
where $N$ is the average population of the reactant region and
\begin{equation}
    \label{fluxDef}
    J = \avg{v_x\,\chi_\text{r}(v_x,\vec q_\bot, \vec v_\bot)}_{\alpha,\text{IC}}
\end{equation}
is the reactive flux out of that region.
Here, $v_x$ denotes the velocity component perpendicular to the DS,
$\vec q_\bot$ the coordinates within the surface and
$\vec v_\bot$ the corresponding velocities.
The characteristic function $\chi_\text{r}(v_x,\vec q_\bot, \vec v_\bot)$ takes the value~1
if the trajectory starting at $x=x^\ddag, v_x, \vec q_\bot, \vec v_\bot$ is reactive,
i.e., moves to products for large times, and~0 otherwise.
Its purpose is to ensure that only reactive trajectories contribute to the reactive flux.
The average in Eq.~\eqref{fluxDef} extends over the realizations, $\alpha$, of the external
noise and over a thermal equilibrium ensemble of initial conditions that are constrained
to lie on the DS.
The latter ensemble is described by a probability density function
\begin{equation}
    \label{pBoltz}
    p(x,v_x,\vec q_\bot, \vec v_\bot) = \delta(x-x^\ddag)\,\exp\left(-\frac{v_x^2}{2\kT}\right)\,
        p_\bot (\vec q_\bot, \vec v_\bot),
\end{equation}
which includes a Boltzmann distribution of the velocities $v_x$ and a Boltzmann distribution
\begin{equation}
    \label{pPerp}
    p_\bot(\vec q_\bot, \vec v_\bot)
        = \frac{1}{Z} \exp\left(-\frac{\vec v_\bot^2/2 + U(x^\ddag,\vec q_\bot)}{\kT}\right)
\end{equation}
of the transverse coordinates and velocities.
The factor~$Z$ in Eq.~\eqref{pPerp} is the partition function of the transverse motion.
It ensures that
\[
    \int d\vec q_\bot\,d\vec v_\bot\, p(\vec q_\bot, \vec v_\bot) = 1.
\]
In Eq.~\eqref{pBoltz} we have used mass-scaled coordinates and we have left out an overall
normalization factor.
In particular, we did not include the Arrhenius factor
\[
    \exp\left(-\frac{\Delta E^\ddag}{\kT}\right)
\]
that includes the activation energy $\Delta E^\ddag$ of the reaction.
The overall normalization of the distribution function is well understood,
and it is irrelevant to the calculation of the transmission factor~\eqref{kappaFlux} below,
on which we will focus in this work.
For simplicity, we can therefore work with the unnormalized distribution
function~\eqref{pBoltz}.

The characteristic function $\chi_\text{r}$ in Eq.~\eqref{fluxDef} encodes the entire
complexity of the reaction dynamics on an anharmonic barrier.
The main task of a reaction rate calculation is to evaluate this function.
In general, this can only be achieved by a numerical simulation.
A simple approximation to this crucial ingredient is provided by TST.
It assumes that no trajectory can cross the DS more than once.
As a consequence, every trajectory that crosses the DS from the reactant
to the product side must be reactive, every trajectory that crosses in the opposite
direction must be nonreactive.
To implement this approximation,
we replace the characteristic function in Eq.~\eqref{fluxDef} by
\begin{equation}
    \label{chiTST}
    \chi^\text{TST} (v_x, \vec q_\bot, \vec v_\bot) =
        \begin{cases}
            1 &:\quad v_x>0, \\
            0 &:\quad v_x<0.
        \end{cases}
\end{equation}
This gives rise to the TST approximation to the rate constant
\begin{equation}
    \label{kTST}
    k^\text{TST} = \frac{\avg{v_x\,\chi^\text{TST}(v_x,\vec q_\bot, \vec v_\bot)}_{\text{IC}}}
                    {N},
\end{equation}
in which the average over the noise $\alpha$ can be suppressed because $\chi^\text{TST}$
does not depend on it.

When the no-recrossing assumption of TST is not satisfied, the approximation~\eqref{kTST}
will overestimate the rate, often by a large factor.
To quantify the effects of non-TST behavior, a transmission factor,
\[
    \kappa = \frac{k}{k^\text{TST}} \le 1,
\]
is introduced that relates the exact rate to the TST approximation.
It can be obtained from the ratio of the flux across the barrier to its TST approximation:
\begin{equation}
    \label{kappaFlux}
    \kappa = \frac{\avg{v_x \chi_\text{r}(v_x,\vec q_\bot, \vec v_\bot)}_{\alpha,\text{IC}}}
        {\avg{v_x \chi^\text{TST}(v_x,\vec q_\bot, \vec v_\bot)}_\text{IC}}.
\end{equation}
To evaluate~\eqref{kappaFlux} numerically, one can randomly sample initial conditions and noise
sequences from the appropriate ensembles, and simulate the behavior of each trajectory until
its energy is so far below the barrier top that it can be regarded as having been thermalized
on either the reactant or the product side of the barrier.
The trajectory can then be classified as reactive or non-reactive depending on what state it reached.
All numerical results presented in this work were obtained in this way.

This algorithm is conceptually straight-forward, but computationally costly.
It would be highly desirable to find a criterion that allows one to identify the reactive
trajectories without having to carry out a numerical simulation.
The following sections will describe the phase space structures that will provide such a criterion.

\section{Time-dependent invariant manifolds}
\label{sec:tdmf}

\subsection{The Langevin model}
\label{sec:lang}

We begin by specifying the model that will be used.
The Langevin equation describes the reduced dynamics of a low-dimensional system coupled
to an external heat bath.\cite{Haenggi90}
It is given by
\begin{equation}
    \label{Langevin}
    \ddot {\vec q} = - \nabla_{\vec q}U(\vec q) - \boldsymbol \Gamma \dot {\vec q} + \vec\xi_\alpha(t),
\end{equation}
where $\vec q$ is an $N$ dimensional vector of mass-scaled coordinates,
$U(\vec q)$ is the potential of mean force,
$\boldsymbol \Gamma$ is a symmetric positive-definite $N\times N$ matrix of damping constants,
and $\vec\xi_\alpha(t)$ is the fluctuating force exerted by the heat bath.
It is connected to the friction matrix $\boldsymbol \Gamma$ by the fluctuation--dissipation
theorem~\cite{Zwanzig01}
\begin{equation}
    \label{flucdis}
    \left\langle \vec \xi_\alpha(t) \vec\xi_\alpha^\text{T}(t')\right\rangle_\alpha =
        2\kT \boldsymbol \Gamma\,\delta(t-t'),
\end{equation}
where $k_{\rm B}$ is the Boltzmann constant and $T$ is the temperature.
Throughout most of this work, we consider a one-dimensional problem in which 
the friction matrix $\boldsymbol \Gamma$ simply reduces to a scalar $\gamma$, 
and the position vector $\vec q$ contains a single coordinate $x$.
If we expand the potential of mean force around its saddle point, we can write it as
\begin{equation}
    \label{Potential1D}
    U(x) = -\tfrac 12 \omb^2 x^2 + \varepsilon\frac{c_3}3 x^3
        + \varepsilon^2\frac{c_4}4 x^4 + \dots.
\end{equation}
where $\varepsilon$ is a formal perturbation parameter that serves only to keep track of
the orders of perturbation theory, and finally will be set to $\varepsilon=1$.
For the mean force itself we write
\begin{equation}
    \label{meanForce}
    -\frac{{\rm d}U}{{\rm d}x} = \omb^2 x + f(x),
\end{equation}
where $f(x)$ denotes the anharmonic parts of the force.

\subsection{Time-dependent transition states}
\label{sec:tdts}

Because the Langevin expression~\eqref{Langevin} is a second order differential equation,
its phase space is two-dimensional, with
coordinates $x$ and $v_x=\dot x$.
As it was observed in Refs.~\onlinecite{Bartsch05b,Bartsch05c,Bartsch06a},
the dynamics of the Langevin equation in the harmonic approximation can be diagonalized
by rewritting it in coordinates $u$ and $s$ given by
\begin{align}
    \label{suTransform}
    u &= \frac{v_{x} - \ls x}{\lu-\ls}, & s &= \frac{-v_{x}+\lu x}{\lu-\ls}, \\
\intertext{or}
    x &= u+s, & v_x &= \lu u + \ls s.
\end{align}
The constants
\begin{equation}
    \label{eval}
    \lambda_\text{s,u} =
        -\frac 12 \left(\gamma \pm \sqrt{\gamma^2+4\omb^2}\right)
\end{equation}
are the eigenvalues that arise in the diagonalization.
They satisfy $\ls<0<\lu$ and
\[
    \lu+\ls=-\gamma, \qquad \lu\ls=-\omb^2.
\]
In the new set of coordinates, the equations of motion read
\begin{align}
    \label{usEqns}
    \dot u &= \lu u + \frac{f(x)}{\lu-\ls}
        + \frac{1}{\lu-\ls}\,\xi_\alpha(t), \nonumber \\
    \dot s &= \ls s - \frac{f(x)}{\lu-\ls}
        - \frac{1}{\lu-\ls}\,\xi_\alpha(t).
\end{align}
These equations decouple in the harmonic approximation, i.e., if $f(x)=0$,
but they are still subject to the time-dependent stochastic driving force $\xi_\alpha(t)$.
This time dependence can be removed
by the coordinate shift
\begin{equation}
    \label{relCoords}
    \Delta u = u - u^\ddag, \qquad \Delta s = s - s^\ddag,
\end{equation}
where
\begin{equation}
    \label{TStraj}
    u^\ddag(t) = \frac{1}{\lu-\ls}\,S[\lu, \xi_\alpha;t], \quad
    s^\ddag(t) = -\frac{1}{\lu-\ls}\,S[\ls, \xi_\alpha;t],
\end{equation}
and the $S$ functionals~\cite{Bartsch05b,Kawai07a} are given by
\begin{equation}
    \label{SDef}
    S_\tau[\mu, g;t] = \begin{cases}
            \displaystyle -\int_t^\infty g(\tau)\,\exp(\mu(t-\tau)) \,d\tau \!\!\!
                & :\; \Re\mu>0, \\[3ex]
            \displaystyle +\int_{-\infty}^t g(\tau)\,\exp(\mu(t-\tau)) \,d\tau \!\!\!
                & :\; \Re\mu<0.
        \end{cases}
\end{equation}
The subscript $\tau$ is used in the $S$ functional to indicate the integration variable.
This subscript will be left out whenever this does not cause any ambiguities.
Similarly, we have for the sake of simplicity not indicated in our notation that
$u^\ddag(t)$ and $s^\ddag(t)$ depend on the realization $\alpha$ of the noise,
although they both obviously do.

The functions $u^\ddag(t)$ and $s^\ddag(t)$
solve the equations of motion in the harmonic limit $f(x)=0$.
They can therefore be regarded as the coordinates of a
special trajectory called the TS~trajectory.
This trajectory is distinguished from all other trajectories that are exposed
to the same noise by the fact that it remains in the vicinity of the saddle point for all times,
whereas a typical trajectory would descend into either the reactant or the product well
both in the remote past and in the distant future.
Accordingly, when using coordinates $\Delta u$ and $\Delta s$, we are
describing a trajectory relative to the TS trajectory,
which acts as a moving coordinate origin.
In what follows, we will refer to $\Delta u$ and $\Delta s$ as relative coordinates
and to the original $u$ and $s$, or $x$ and $v_x$ as space fixed coordinates.

The equations of motion in relative coordinates are
\begin{subequations}
\label{usDeltaEqnsOrig}
\begin{align}
    \Delta\dot u &= \lu \Delta u + \frac{f(x)}{\lu-\ls},
        \label{uDeltaOrig} \\
    \Delta \dot s &= \ls \Delta s - \frac{f(x)}{\lu-\ls}.
        \label{sDeltaOrig}
\end{align}
\end{subequations}
At first sight, it appears that the time-dependent and stochastic shift~\eqref{relCoords}
has removed both the time-dependence and the dependence of the realization $\alpha$ of the noise.
However, this is only true in the harmonic approximation.
If we express the position
coordinate $x$ in terms of the relative coordinates $\Delta u$ and $\Delta s$ as
\begin{equation}
    \label{xInRel}
    x = x^\ddag + \Delta u+\Delta s,
\end{equation}
with $x^\ddag = u^\ddag + s^\ddag$,  Eq.~\eqref{usDeltaEqnsOrig} turns into
\begin{subequations}
\label{usDeltaEqnsGeneral}
\begin{align}
    \Delta\dot u &= \lu \Delta u + \frac{f(x^\ddag+\Delta u + \Delta s)}{\lu-\ls},
        \label{uDeltaGeneral} \\
    \Delta \dot s &= \ls \Delta s - \frac{f(x^\ddag+\Delta u + \Delta s)}{\lu-\ls}.
        \label{sDeltaGeneral}
\end{align}
\end{subequations}
The position $x^\ddag(t)$ of the TS~trajectory represents a time-dependent stochastic
driving in these equations of motion.
Nevertheless, the coordinate shift has removed the stochastic driving from the leading-order
terms in~\eqref{usDeltaEqnsGeneral} and pushed it into the anharmonic perturbation.

The description of the geometric phase space structure in the vicinity of the saddle point
is most easily done if one starts from the harmonic limit.
A full discussion can be found in Refs.~\onlinecite{Bartsch05b,Bartsch05c}.
The equations of motion~\eqref{usDeltaEqnsOrig} decouple and become time independent
when $f(x)=0$, and they can then be easily solved by writing
\begin{align}
    \label{deltaSolHarm}
    \Delta u(t) &= \Delta u(0)\,e^{\lu t}, \nonumber \\
    \Delta s(t) &= \Delta s(0)\,e^{\ls t}.
\end{align}
Since $\lu>0$ and $\ls<0$, the coordinate $\Delta u$ grows exponentially in time,
whereas $\Delta s$ shrinks.
Therefore, $\Delta u$ and $\Delta s$ correspond to unstable and stable directions in phase space,
respectively.
In particular, the lines $\Delta u=0$ and $\Delta s=0$ are invariant under the dynamics.
A trajectory that starts on the line $\Delta u=0$ will asymptotically approach the origin
as $t\to\infty$; this line is called the stable manifold of the origin.
A trajectory on the line $\Delta s=0$ will move away from the origin as $t\to\infty$,
but it will approach the origin as $t\to-\infty$;
this line is called the unstable manifold of the origin.
\begin{figure}
\includegraphics[]{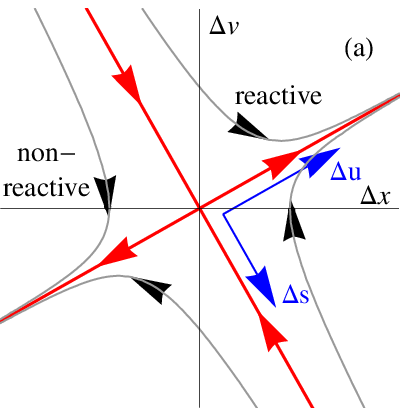}\hfill
\includegraphics[]{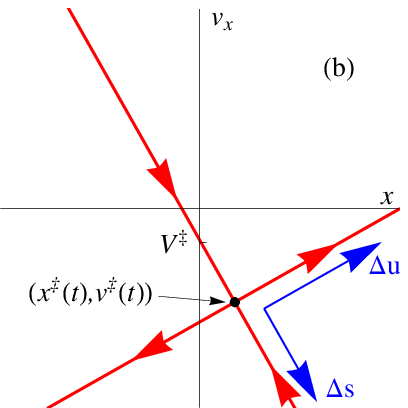}\\[3ex]
\includegraphics[]{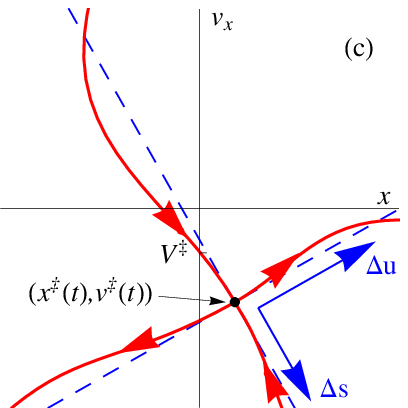}\hfill
\includegraphics[]{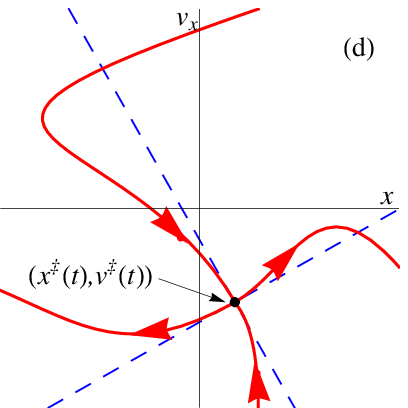}
\caption{Phase space view of the time-dependent invariant manifolds of the Langevin equation.
  (a) Invariant manifolds are time-independent in the harmonic approximation and in
      relative coordinates.
  (b) In space-fixed coordinates, the invariant manifolds are attached to the TS trajectory
      and move through phase space with it.
  (c) Anharmonic coupling deforms the manifolds. Both their position and their shape are
      stochastically time dependent.
  (d) Invariant manifolds can deviate strongly from the harmonic approximation if the
      anharmonicities are strong.
  }
\label{fig:mf}
\end{figure}

The stable and unstable manifolds of the origin, together with several typical trajectories
in relative coordinates, are shown in Fig.~\ref{fig:mf}a.
The invariant manifolds separate trajectories with different qualitative behavior.
Trajectories above the stable manifold, i.e., with larger relative velocity,
move to the product side of the barrier for asymptotically long times,
whereas trajectories below the stable manifold move to the reactant side.
Similarly, trajectories above the unstable manifold come from the reactant side in the distant past,
whereas trajectories below the unstable manifold come from the product side.

For a reaction rate calculation we need to ascertain
whether a trajectory will turn into reactants or products in the future.
In our approach this sentence is
rephrased into the condition: We need to decide whether a trajectory lies above
or below the stable manifold.
In other words,
the stable manifold encodes the information about the reaction dynamics that is most relevant to us.
We will therefore focus on the stable manifold in what follows, largely ignoring the unstable manifold.

We can return to space fixed coordinates by undoing the time dependent shift~\eqref{relCoords}.
After the shift, the stable and unstable manifolds are not attached to the origin of the
coordinate system any more, but instead to the TS trajectory as a moving origin,
as shown in Fig.~\ref{fig:mf}(b).
Since the TS trajectory is time dependent, the manifolds will move through phase space with it.
Nevertheless, they still separate trajectories with different asymptotic behaviors.
Given a trajectory with a given initial condition at a certain time,
it can be classified as reactive or non-reactive by knowing the instantaneous position
of the stable manifold at that time.
Through the TS trajectory, that instantaneous position will depend on the realization of the noise.

It is clear from Fig.~\ref{fig:mf}(b) that at any time and for any realization of the noise
the stable manifold intersects the axis $x=0$ at a point with a velocity $V^\ddag$.
Trajectories with initial positions $x=0$ and initial velocities $v_x>V^\ddag$ are reactive,
while trajectories with initial velocities $v_x<V^\ddag$ are not.
The critical velocity $V^\ddag$ depends on time and on the realization of the noise.
For the harmonic approximation, it was shown in Ref.~\onlinecite{Bartsch08},
and it will be rederived below, that
\begin{equation}
    \label{V0}
    V^\ddag \equiv V^\ddag_0 = (\lu-\ls)u^\ddag(0).
\end{equation}
Since the critical velocity characterizes reactive trajectories,
the transmission factor~\eqref{kappaFlux} can be expressed in terms of $V^\ddag$
(see Ref.~\onlinecite{Bartsch08} and Section~\ref{sec:rate} below).

This picture of the invariant manifolds was introduced in Refs.~\onlinecite{Bartsch05b,Bartsch05c}
and applied to rate calculations in Refs.~\onlinecite{Bartsch06a,Bartsch08}.
The main purpose of the present work is to explore how this picture changes when
anharmonicities of the barrier potential are taken into account.
In this case the equations of motion~\eqref{usDeltaEqnsOrig} are coupled in a nonlinear
time-dependent way, and they cannot be solved easily.
However, as long as the coupling is sufficiently weak, it can be expected to find
a TS trajectory and with its associated stable and unstable manifolds that are close to
those in the harmonic approximation.
Indeed, there are general theorems in the theory of stochastic dynamical systems\cite{Arnold98}
that guarantee the persistence of these structures.
As shown in Fig.~\ref{fig:mf}(c), the invariant manifolds in an anharmonic system will be
tangent to their harmonic approximations at the TS trajectory,
but they will not be straight lines anymore.
Because the coupling term in~\eqref{usDeltaEqnsOrig} is stochastically time dependent,
the shapes of the invariant manifolds as well as their positions in phase space
depend on time and on the realization of the noise.

The intersection of the stable manifold with the axis $x=0$ will give rise to a critical
velocity $V^\ddag$ such that trajectories with initial velocities larger than $V^\ddag$
will be reactive, those with smaller initial velocities will not.
The critical velocity can therefore be used in a rate calculation in an
anharmonic system just as it can in the harmonic approximation,
though its value will be different from~\eqref{V0}.
A method to calculate the critical velocity will be developed in Section~\ref{sec:perturb1D}.

In general it cannot be guaranteed that there will only be a single intersection
between the stable manifold and the axis $x=0$.
In fact, if the reaction potential has wells on the reactant and/or product
side of the barrier, it is likely that there will be further intersections, as illustrated
in Fig.~\ref{fig:mf}(d).
If a trajectory on the stable manifold is followed backwards in time,
it will descend from the barrier, settling in one of the wells for some time.
If it is followed for long enough, it will eventually cross the barrier again
into the other well.
In doing so, it must cross the line $x=0$ again, and thus give rise to additional
intersections between the stable manifold and that line.
However, as these additional intersections stem from previous barrier crossings,
they must be neglected in the rate calculation.
Only for extremely strong nonlinearities additional intersections that are not
separated by periods in which the trajectory was equilibrated in one of the wells
will be found.
We will neglect that possibility in what follows.
Instead, we will apply perturbation theory to calculate a value for the critical
velocity that reduces to its harmonic approximation in the appropriate limit.

The TS trajectory~\eqref{TStraj} solves the equations of motion~\eqref{usEqns}
in the harmonic limit, but not in the presence of anharmonic coupling.
Strictly speaking, therefore, Eq.~\eqref{TStraj} does not define a TS~trajectory
on an anharmonic potential.
Such a trajectory could be obtained by a perturbative expansion similar to the one
to be developed in Section~\ref{sec:perturb1D}.
For our purposes, however, this will not be necessary.
The harmonic TS~trajectory forms a suitable basis for the perturbation theory.
We will therefore use the notation $u^\ddag$, $s^\ddag$ and $x^\ddag$
exclusively to denote the harmonic approximation to the TS~trajectory.

\section{Perturbative calculation of the stable manifold}
\label{sec:perturb1D}

The critical velocity is defined by the intersection of the line $x=0$
with the stable manifold of the TS~trajectory.
The stable manifold contains all those trajectories that approach the
TS~trajectory as $t\to\infty$.
They remain bounded for large times.
Solutions to the equations of motion~\eqref{usDeltaEqnsGeneral} that satisfy this
boundary condition at large time lie on the stable manifold.

Equation~\eqref{uDeltaGeneral} can be formally solved in terms of the $S$~functional
(\ref{SDef}) as
\[
    \Delta u(t) = C e^{\lu t} + \frac{1}{\lu-\ls}\,S[\lu,f(x^\ddag+\Delta u + \Delta s); t].
\]
Notice that this is only a formal solution due to the presence of
the unknown function $\Delta u$ in the r.h.s. of the equation.
Furthermore, the $S$~functional is undefined for most trajectories,
only existing for the trajectories that remain bounded in the remote future.
However, these are precisely the trajectories we are interested in.
For consistency, we must then set $C=0$, just as was done in
Refs.~\onlinecite{Bartsch05b,Bartsch05c} in the construction of the TS~trajectory.
A trajectory on the unstable manifold therefore satisfies the integral equation
\begin{equation}
    \label{uEqInt}
    \Delta u(t) =  \frac{1}{\lu-\ls}\,S[\lu,f(x^\ddag+\Delta u + \Delta s); t].
\end{equation}
This expression automatically incorporates the boundary condition
at $t\to\infty$ that we wish to impose.

For the stable component, we might be tempted to use the analogous formal solution
\[
    \Delta s(t) = C e^{\ls t} - \frac{1}{\lu-\ls}\,S[\ls,f(x^\ddag+\Delta u + \Delta s); t].
\]
However, the $S$ functional for a negative eigenvalue depends on the infinite past of
its argument and is well defined only for trajectories that remain bounded in the past.
Most trajectories on the stable manifold, except for the TS~trajectory itself,
will not satisfy this condition.
This difficulty can be circumvented by using the modified $S$~functional
\begin{equation}
    \label{SmodDef}
    \bar S_\tau[\mu,g;t] = \int_0^t g(\tau) e^{\mu(t-\tau)}\,d\tau
\end{equation}
that is well defined for all values of $\mu$.
It satisfies the differential equation
\[
    \frac{d}{dt} \bar S[\mu,g;t] = \mu\, \bar S[\mu,g;t] + g(t)
\]
and the initial condition $\bar S[\mu,g;0]=0$.
With this functional, a formal solution to the equation of motion~\eqref{sDeltaGeneral}
can be written as
\begin{equation}
    \label{sEqInt}
    \Delta s(t) = \Delta s(0) e^{\ls t}
            - \frac{1}{\lu-\ls}\,\bar S[\ls,f(x^\ddag+\Delta u + \Delta s); t].
\end{equation}
Note that this integral equation does not impose any boundary condition
on the function $\Delta s$, thus leaving free choice of
the initial condition~$\Delta s(0)$.

The critical velocity $V^\ddag$ is determined by the condition that the trajectory with
initial conditions $x(0)=0$ and $v(0)=V^\ddag$ satisfies the integral
equations~\eqref{uEqInt} and~\eqref{sEqInt}.
The first one of these conditions can be rewritten as
\[
    \Delta s(0) = -x^\ddag(0) - \Delta u(0),
\]
such that the initial condition for $\Delta s$, which is needed in Eq.~\eqref{sEqInt},
is known once the initial value of $\Delta u$ has been determined from Eq.~\eqref{uEqInt}.
The critical velocity can then be obtained from
\begin{align}
    \label{VCritU}
    V^\ddag &= v(0) = \lu u(0) + \ls s(0) \nonumber \\
        &= (\lu-\ls) u(0) \qquad\qquad\text{as $x(0)=u(0)+s(0)=0$}\nonumber\\
        &= (\lu-\ls)[u^\ddag(0) + \Delta u(0)].
\end{align}

In the harmonic approximation the trajectory that starts in the DS $x=0$
and lies in the stable manifold is given by
\begin{equation}
    \label{DeltaHarm}
    \Delta u_0(t)= 0 \qquad \text{and} \qquad
    \Delta s_0(t) = -x^\ddag(0) e^{\ls t}.
\end{equation}
For this case, Eq.~\eqref{VCritU} leads back to the result~\eqref{V0}
\[
    V^\ddag_0=(\lu-\ls)u^\ddag(0).
\]
When the solution~\eqref{DeltaHarm} is substituted into the integral equations~\eqref{uEqInt}
or~\eqref{sEqInt}, the coordinate $x=x^\ddag+\Delta u + \Delta s$ is replaced by
\begin{equation}
    \label{XDef}
    X (t) = x^\ddag(t)-e^{\ls t}x^\ddag(0).
\end{equation}
This function represents the harmonic approximation to the coordinate $x(t)$
of the trajectory under study.
Moreover, it constitutes a suitable basis of the perturbative expansion.

The leading-order correction to the critical velocity can be obtained from~\eqref{uEqInt}
as
\[
    \Delta u_\text{lead}(t) = \frac{1}{\lu-\ls} S[\lu,f(X); t],
\]
from which it follows that
\begin{align}
    \label{VCritLead}
    V^\ddag_\text{lead}
        &= S[\lu, f(X);0].
\end{align}

To obtain higher-order corrections to the critical velocity in a systematic manner,
we introduce the expansions
\begin{alignat*}{5}
    V^\ddag &= V^\ddag_0 & + & \varepsilon\, V^\ddag_1 &&+ \varepsilon^2\, V^\ddag_2 &&+ \dots \\
    \Delta u &= && \varepsilon\, \Delta u_1 && + \varepsilon^2\, \Delta u_2 &&+ \dots\\
    \Delta s &= -x^\ddag & + & \varepsilon \,\Delta s_1 && + \varepsilon^2 \,\Delta s_2 && + \dots
\end{alignat*}
We will write
\begin{equation}
    \label{DeltaXDef}
    \Delta x_k=\Delta u_k + \Delta s_k \qquad \text{for $k\ge 1$.}
\end{equation}
Expand the anharmonic term as
\begin{equation}
    \label{forceExp}
    f(X + \varepsilon\,\Delta x_1 + \varepsilon^2\,\Delta x_2 + \dots ) =
    \varepsilon\,f_1 + \varepsilon^2\,f_2 + \dots,
\end{equation}
where terms in the r.h.s. depend on the $\Delta x_j$.
Since $f$ is assumed to have an overall order $\varepsilon$ or higher,
the calculation of the term $f_k$ requires only the knowledge of $\Delta x_j$ for $j<k$.
Equations~\eqref{uEqInt}, \eqref{sEqInt} and~\eqref{DeltaXDef} then yield the
recurrence relations
\begin{align}
    \label{perturbRecur}
    \Delta u_k(t) &= \frac{1}{\lu-\ls}\,S[\lu,f_k;t], \nonumber \\
    \Delta s_k(t) &= -\Delta u_k(0) e^{\ls t} - \frac{1}{\lu-\ls}\bar S[\ls,f_k;t],
        \nonumber \\
    \Delta x_k(t) &= \Delta u_k(t) + \Delta s_k(t),
\end{align}
from which it can be finally obtained
\begin{equation}
    \label{VCritCorr}
    V^\ddag_k = (\lu-\ls)\Delta u_k(0).
\end{equation}
The recursion relations~\eqref{perturbRecur} can be successively evaluated as written for
$k=1,2,\dots$ up to any desired order.

For example, for the anharmonic force corresponding to the generic one-dimensional
potential~\eqref{Potential1D} with only cubic and quartic terms,
expansion~\eqref{forceExp} gives
\begin{align*}
    f_1 &= -c_3 X^2, \\
    f_2 &= -c_4 X^3 - 2c_3 X\,\Delta x_1.
\end{align*}
It is then obtained
\begin{widetext}
\begin{align*}
    \Delta u_1(t) &= -\frac{c_3}{\lu-\ls}\,S[\lu, X^2;t], \\[1ex]
    \Delta s_1(t) &= \frac{c_3}{\lu-\ls}\left( S[\lu,X^2;0] e^{\ls t} +
        \bar S[\ls,X^2;t]\right), \\[1ex]
    \Delta x_1(t) &= \frac{c_3}{\lu-\ls}\left( S[\lu,X^2;0] e^{\ls t}
        -S[\lu,X^2;t]+\bar S[\ls,X^2;t]\right) ,\\[1ex]
    \Delta u_2(t) &= -\frac{1}{\lu-\ls} \,
        S[\lu, 2 c_3 X\,\Delta x_1 + c_4 X^3; t] \\
    &= -\frac{c_4}{\lu-\ls}\,S[\lu,X^3; t]
     - \frac{2c_3^2}{(\lu-\ls)^2} S_\tau\left[\lu, X(\tau)
        \left(  e^{\ls \tau} S[\lu,X^2;0] - S[\lu,X^2;\tau]
            + \bar S[\ls,X^2;\tau]\right); t \right].
\end{align*}
From~\eqref{VCritCorr} we have that
\begin{equation}
    \label{V1general}
    V^\ddag_1 = -c_3\, S[\lu, X^2;0],
\end{equation}
in agreement with Eq.~\eqref{VCritLead}, and
\begin{align}
    \label{V2general}
    V^\ddag_2 ={}& -c_4\,S[\lu,X^3; 0]
     - \frac{2c_3^2}{\lu-\ls} S_\tau\left[\lu, X(\tau) \left(
            e^{\ls \tau} S[\lu,X^2;0] - S[\lu,X^2;\tau]
            + \bar S[\ls,X^2;\tau]\right); 0 \right].
\end{align}
\end{widetext}
Not surprisingly,
the corrections~\eqref{V1general} and~\eqref{V2general} depend, through the function $X$,
on the realization of the noise.
This dependence reflects the fact that on an anharmonic potential not only the position,
but also the shape of the invariant manifolds, are stochastically time dependent.

\begin{figure}
\includegraphics{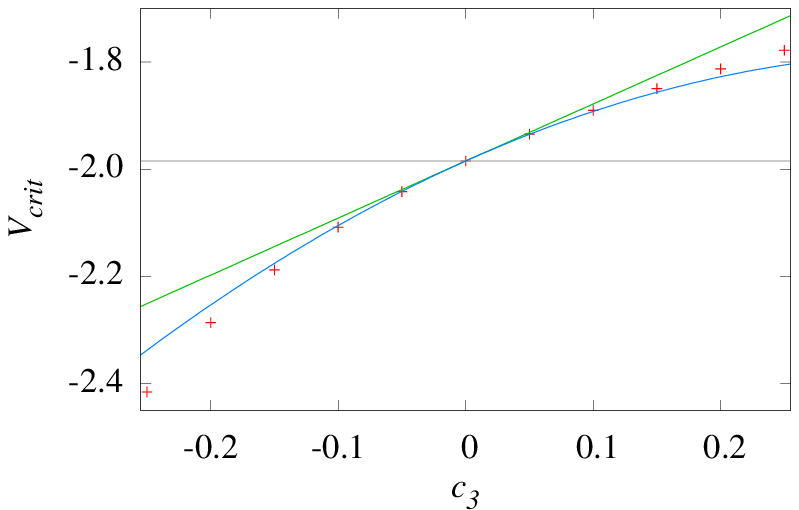}
 \caption{Critical velocity for a realization of the noise for a one-dimensional barrier
   with cubic anharmonicity, $c_3$, for $\omb=1$, $\gamma=2$, $\kT=1$: \\
   Numerical simulation results (red crosses), harmonic approximation~\eqref{V0} (gray horizontal line),
   perturbative results to first-order~\eqref{V0}+\eqref{V1general} (green straight line) and
   second-order~\eqref{V0}+\eqref{V1general}+\eqref{V2general} (blue line).}
 \label{fig:VCritCubic}
\end{figure}

We also calculated the critical velocity numerically for a given realization of the noise.
To this end, an ensemble of trajectories starting on the DS
was propagated numerically.
By recording which trajectories were reactive and which were not,
the value of the critical velocity could be bracketed with high accuracy.
For one fixed realization and for a potential with only a cubic anharmonic term,
the perturbative expansion is compared to numerical results in Fig.~\ref{fig:VCritCubic}.
There is good agreement between perturbative and numerical results.
Similar figures are obtained for other realizations of the noise,
thus leading to the same conclusion.
Obviously, the size of the first and second order corrections,
as well as that of the higher order terms that are omitted,
varies among different realizations.

\begin{figure}
\includegraphics{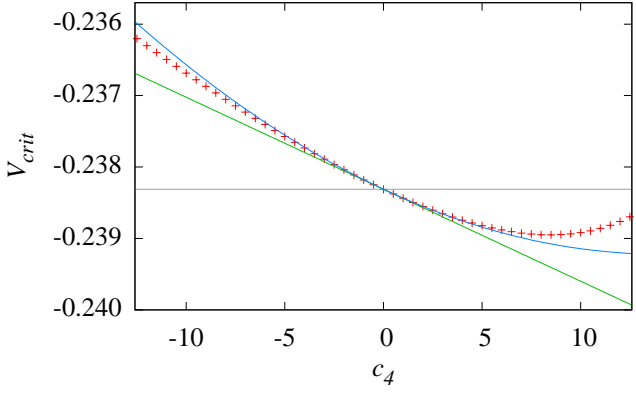}
 \caption{Critical velocity for one realization of the noise for a one-dimensional barrier
   with quartic anharmonicity, $c_4$, for $\omb=1$, $\gamma=2.5$, $\kT=1$: \\
   Numerical simulation results (red crosses), harmonic approximation~\eqref{V0} (gray horizontal line),
   perturbative results to first-order~\eqref{V0}+\eqref{V2quartic} (green straight line) and
   second-order~\eqref{V0}+\eqref{V2quartic}+\eqref{V4quartic} (blue line).}
 \label{fig:VCritQuartic}
\end{figure}

In the special case that the anharmonic potential contains only a quartic term,
the perturbation expansion results as an expansion in powers of $\varepsilon^2$,
with the odd orders terms null.
For the first two non-zero corrections, a similar calculation shows that
\begin{equation}
    \label{V2quartic}
    V^\ddag_2 = -c_4\, S[\lu, X^3;0],
\end{equation}
which is again consistent with Eq.~\eqref{VCritLead}, and
\begin{widetext}
\begin{align}
    \label{V4quartic}
    V^\ddag_4 ={}
     - \frac{3c_4^2}{\lu-\ls} S_\tau\left[\lu, X^2(\tau) \left(
            e^{\ls \tau} S[\lu,X^3;0] - S[\lu,X^3;\tau]
            + \bar S[\ls,X^3;\tau]\right); 0 \right].
\end{align}
\end{widetext}
A comparison of the perturbative corrections~\eqref{V2quartic} and~\eqref{V4quartic}
with numerical results is shown in Fig.~\ref{fig:VCritQuartic}.
Again, this comparison confirms the accuracy of the perturbative results.

The function $X$ introduced in Eq.~\eqref{XDef} plays a special role in the perturbation
expansion because it represents the unperturbed trajectory.
To obtain a different perspective of this function, note that the critical velocity should
depend only on the behavior of the stochastic force $\xi_\alpha(t)$ for $t\ge 0$,
but not on the driving at earlier times:
Once the initial conditions of a trajectory at $t=0$ are given,
its future fate can only depend on the future noise.
The separatrix between reactive and nonreactive trajectories must therefore also be
determined by only the future noise.
Yet the perturbation term in~\eqref{usDeltaEqnsGeneral} depends, via $x^\ddag(t)$,
on $s^\ddag(t)$, which is given by past noise.

If we split up the integration range of the $S$ functional, we find that for $t\ge 0$
\[
    s^\ddag(t) = e^{\ls t} s^\ddag(0) +
        \int_0^t e^{\ls(t-\tau)}\xi_\alpha(\tau)\,d\tau.
\]
The integral in this expression depends only on noise for $t\ge 0$.
The term including $s^\ddag(0)$ contains all the dependence on the past,
and it drops out when we form $X(t)$.
The variable $X$ is the simplest modification of $x^\ddag$ in which the dependence
on the past has been removed.

\section{Corrections to the reaction rates}
\label{sec:rate}

\subsection{General rate expressions}
\label{sec:rateGen}

In a one-dimensional model, the characteristic function $\chi_\text{r}$
can be expressed in terms of the critical velocity as
\begin{equation}
    \label{chiVcrit1D}
    \chi_\text{r} (v_x) =
        \begin{cases}
            1 &:\quad v_x>V^\ddag, \\
            0 &:\quad v_x<V^\ddag.
        \end{cases}
\end{equation}
In contrast to the TST~approximation~\eqref{chiTST}, and in spite of its simplicity,
the expression~\eqref{chiVcrit1D} is exact.
It allows to evaluate the average over initial conditions in
Eq.~\eqref{kappaFlux}---the factor $p_\bot$ in Eq.~\eqref{pBoltz} being absent in one
dimension---to find
\begin{equation}
    \label{kappaVel}
    \kappa = \avg{ \exp\left(-\frac{V^{\ddag2}}{2\kT}\right) }_\alpha,
\end{equation}
where only the average over the noise remains.
This expression was derived in Ref.~\onlinecite{Bartsch08} for a harmonic barrier.
It is now clear that the same expression holds also for anharmonic potentials if the
critical velocity $V^\ddag$ is suitably modified.
Remarkably, no anharmonic corrections arise in the rate expression~\eqref{kappaVel}.

If we have a perturbative expansion
\begin{equation}
    \label{Vexpand}
    V^\ddag = V_0^\ddag + \varepsilon V_1^\ddag + \varepsilon^2 V_2^\ddag + \dots,
\end{equation}
we can substitute into~\eqref{kappaVel} and expand the exponential to obtain a series
of rate corrections
\begin{equation}
    \label{kappaExpand}
    \kappa = \kappa_0 + \varepsilon \kappa_1 + \varepsilon^2 \kappa_2 + \dots,
\end{equation}
where
\begin{subequations}
\label{kappaAvg}
\begin{align}
    \kappa_0 &= \avg{ P }_\alpha, \label{kappaAvg0} \\
    \kappa_1 &= -\frac{1}{\kT} \avg{ P V_0^\ddag V_1^\ddag}_\alpha, \label{kappaAvg1}\\
    \kappa_2 &= \frac{1}{2(\kT)^2} \avg{P V_0^{\ddag2} V_1^{\ddag2}}_\alpha
        - \frac{1}{\kT} \avg{P V_0^\ddag V_2^\ddag}_\alpha \nonumber \\
        & \quad - \frac{1}{2\kT} \avg{P V_1^{\ddag2}}_\alpha \label{kappaAvg2}
\end{align}
\end{subequations}
with the abbreviation
\begin{equation}
    \label{EDef}
    P = \exp\left(-\frac{V_0^{\ddag2}}{2\kT}\right)
        = \exp\left(-\frac{(\lu-\ls)^2\,u^{\ddag2}(0)} {2\kT}\right).
\end{equation}
We will now address
the problem of evaluating the noise averages in Eq.~\eqref{kappaAvg}.

\subsection{Distorted correlation functions}
\label{sec:moments}

The corrections to the critical velocity that appear in the averages~\eqref{kappaAvg}
are expressed in terms of the function $X(t)$, which is in turn given in terms of the
components $u^\ddag(t)$ and $s^\ddag(t)$ of the TS~trajectory.
They are Gaussian random variables whose correlation functions were evaluated in
Ref.~\onlinecite{Bartsch05c}. In the current notation and with
\begin{equation}
    \label{sigmaDef}
    \sigma^2 = \frac{\kT \gamma}{|\ls|(\lu-\ls)^2},
\end{equation}
they read, for $t\ge 0$, as
\begin{subequations}
\label{suCorrelations}
\begin{align}
    \avg{s^\ddag(t) s^\ddag(0)}_\alpha &= \sigma^2 e^{\ls t}, \\
    \avg{u^\ddag(t) u^\ddag(0)}_\alpha &= -\frac{\ls}{\lu} \sigma^2 e^{-\lu t}, \\
    \avg{u^\ddag(t) s^\ddag(0)}_\alpha &= 0, \\
    \avg{s^\ddag(t) u^\ddag(0)}_\alpha &= \frac{2\ls}{\lu+\ls}\,\sigma^2
        \left(e^{-\lu t} - e^{\ls t}\right).
\end{align}
\end{subequations}

To evaluate the corrections~\eqref{kappaAvg} to the reaction rate,
we need to calculate noise averages of the form $\avg{P (\dots)}_\alpha$,
where $(\dots)$ indicates some expression in the functions $u^\ddag(t)$
and $s^\ddag(t)$.
We will therefore assume that the expression $(\dots)$ can be written as a
function of finitely many random variables $\vec z = (z_1, \dots, z_n)$ that follow a
multidimensional Gaussian distribution with zero mean and covariance matrix $\Sigma$,
i.e., the matrix elements of $\Sigma$ are $\sigma_{ij} = \avg{z_iz_j}_\alpha$.
As the first component we include the variable $z_1=u^\ddag(0)$,
which plays a special role because it occurs in the factor~$P$ in Eq.~\eqref{EDef}.

Using~\eqref{V0} and setting $\rho=(\lu-\ls)^2/\kT$, we can write
\begin{align}
\avg {P (\dots)}_\alpha
    &=
    \frac{1}{\sqrt{(2\pi)^n \det\Sigma}} \int d^n z\,
        e^{-\vec z^\text{T} \Sigma^{-1} \vec z/2}
        e^{-\rho z_1^2/2} (\dots) \nonumber \\
    &=\frac{1}{\sqrt{(2\pi)^n \det\Sigma}} \int d^n z\,
        e^{-\vec z^\text{T} (\Sigma^{-1}+\rho J) \vec z/2}
        (\dots) \nonumber \\
    &= \sqrt{\frac{\det \Sigma_0}{\det\Sigma}} \,
        \frac{1}{\sqrt{(2\pi)^n \det\Sigma_0}} \int d^n z\,
        e^{-\vec z^\text{T} \Sigma_0^{-1} \vec z/2}
        (\dots),  \nonumber \\
    &= \sqrt{\frac{\det \Sigma_0}{\det\Sigma}}\, \avg{...}_0,
    \label{modGaussian}
\end{align}
where we have introduced the matrix
\[
    J = \begin{pmatrix}
        1 & 0 & 0 & \dots \\
        0 & 0 & 0 & \dots \\
        0 & 0 & 0 & \dots \\
        \vdots & \vdots & \vdots & \ddots
    \end{pmatrix}
\]
and we have used $\avg{...}_0$ to denote an average over a multidimensional
Gaussian distribution with the modified covariance matrix $\Sigma_0$ given by
\[
    \Sigma_0^{-1} = \Sigma^{-1} + \rho J.
\]

From the observation
\[
    \Sigma J = \begin{pmatrix}
                \sigma_{11} & 0 & \dots & 0 \\
                \sigma_{21} & 0 & & 0 \\
                \vdots & & \ddots & \vdots \\
                \sigma_{n1} & 0 & \dots & 0
            \end{pmatrix}
\]
we obtain $(\Sigma J)^2 = \sigma_{11} \Sigma J$. It is then easy to check that
\[
    \left(\Sigma-\frac{\rho}{1+\rho\sigma_{11}}\,\Sigma J \Sigma\right)
    \left(\Sigma^{-1}+ \rho J\right)
    = I,
\]
the identity matrix. Therefore
\begin{align}
    \label{modSigma}
    \Sigma_0 &= \Sigma-\frac{\rho}{1+\rho\sigma_{11}}\,\Sigma J \Sigma \nonumber\\
        &= \Sigma + \frac{\lu}{\ls}\rho \Sigma J \Sigma,
\end{align}
where in the last step we have used the value given in Eq.~\eqref{suCorrelations}
for $\sigma_{11} = \avg{u^{\ddag2} (0)}_\alpha$.

Furthermore,
\[
    \Sigma_0 \Sigma^{-1} = I - \frac{\rho}{1+\rho \sigma_{11}} \Sigma J
\]
is a lower triangular matrix whose diagonal elements,
except for the $(1,1)$~element,
are all equal to 1. This observation makes it easy to evaluate
\begin{align}
    \label{detSdetS0}
    \frac{\det\Sigma_0}{\det\Sigma} &=
        \det\left(I - \frac{\rho}{1+\rho \sigma_{11}}\Sigma J\right) \nonumber \\
        &= 1 - \frac{\rho \sigma_{11}}{1+\rho \sigma_{11}} \nonumber \\
        &= -\frac{\lu}{\ls} = \frac{\lu^2}{\omb^2},
\end{align}
where Eq.~\eqref{suCorrelations} has again be used.

Substituting Eq.~\eqref{detSdetS0} in Eq.~\eqref{modGaussian}, we
finally find
\begin{equation}
    \label{avgMod}
    \avg{P (\dots)}_\alpha = \frac{\lu}{\omb}\,\avg{\dots}_0.
\end{equation}
For the components of the modified covariance matrix~\eqref{modSigma} we find
\begin{equation}
    \label{modSigmaComp}
    \avg{z_i z_j}_0 = \avg{z_i z_j}_\alpha
        + \frac{\lu}{\ls}\rho \avg{u^\ddag(0) z_i}_\alpha \avg{u^\ddag(0) z_j}_\alpha,
\end{equation}
which allows to obtain the moments of the distorted Gaussian distribution once the
moments of the original Gaussian are known.
In particular, $\avg{z_i z_j}_0 = \avg{z_i z_j}_\alpha$ if either $z_i$ or $z_j$
are uncorrelated with $u^\ddag(0)$.

Once the second moments of the distorted Gaussian distribution,
i.e., the matrix elements of $\Sigma_0$, are known,
Isserlis' theorem~\cite{Isserlis16,Isserlis18} can be used to express
higher-order moments in terms of second moments, e.g.
\begin{align*}
    &\quad\avg{z_1 z_2 z_3 z_4}_0 \\
    &= \avg{z_1 z_2}_0 \avg{z_3z_4}_0
        + \avg{z_1 z_3}_0 \avg{z_2z_4}_0 + \avg{z_1 z_4}_0 \avg{z_2 z_3}_0.
\end{align*}
This expression contains a sum over all possible pairings of the four factors.
Other even-order moments can be evaluated in a similar way,
and the odd-order moments are zero.
In this way, the modified averages of arbitrary polynomials can be calculated.

The moments that will be required in the rate calculation can be obtained from these results;
they are
\begin{widetext}
\begin{subequations}
\label{XMoments}
\begin{align}
    \frac{\avg{u^\ddag(0) X(t)}_0}{\sigma^2} &=
         (1-\beta_\text{u}) \left(e^{-\lu t} - e^{\ls t}\right), \\
    \frac{\avg{X(t)X(t')}_0}{\sigma^2} &=
        (1-\beta_\text{s}) e^{\ls|t-t'|}  - \frac{\ls}{\lu}(1-\beta_\text{u}) e^{-\lu|t-t'|}
        + \left(1-2\beta_\text{s} + \frac{\ls}{\lu}\right) e^{-\lu(t+t')}
        + (1-\beta_\text{u}) \left(e^{-\lu t + \ls t'} + e^{-\lu t' + \ls t}\right),
\end{align}
\end{subequations}
\end{widetext}
with
\begin{equation*}
    \beta_\text{u} = \frac{2\lu}{\lu+\ls}, \qquad
    \beta_\text{s} = \frac{2\ls}{\lu+\ls}.
\end{equation*}

\subsection{Results for the one-dimensional potential}
\label{sec:results1d}

With the help of Eq.~\eqref{avgMod} the leading term in the transmission
factor~\eqref{kappaAvg0} can be evaluated, giving
\begin{equation}
    \label{kappaKramers}
    \kappa_0 = \frac{\lu}{\omb}.
\end{equation}
This is the famous Kramers result for the transmission factor.\cite{Haenggi90}

The perturbation expansion is set up in such a way that effectively the noise carries a
factor of $\varepsilon$.
The critical velocity $V_0^\ddag$ is linear in the noise.
If $V_1^\ddag$ is one order $\varepsilon$
higher,
it must be quadratic in the noise, and $V_2^\ddag$ cubic. Consequently,
\[
    \kappa_1 = -\frac{1}{\kT}\,\frac{\lu}{\omb}
        \avg{V_0^\ddag V_1^\ddag}_0 =0
\]
is a third-order moment of the noise and must  vanish.
Similarly, all odd-order corrections to the transmission factor must be zero.
According to the fluctuation-dissipation theorem~\eqref{flucdis},
the noise carries a factor $\sqrt{\kT}$, so that a perturbative expansion in powers
of $\varepsilon$ corresponds to an expansion in powers of $\sqrt{\kT}$.
By contrast, Eq.~\eqref{kappaExpand} is an
expansion of the transmission factor in powers of $\kT$ because it has only even-order terms.

The simplest rate correction can therefore be obtained from a quartic perturbation in the potential.
We set $c_3=0$, which makes $V_1^\ddag=0$, and calculate the rate correction that is linear in $c_4$.
Substituting Eqns.~\eqref{V0} and~\eqref{V2quartic} into~\eqref{kappaAvg1}, it is found that
\begin{equation}
    \label{kappa2First}
    \kappa_2^{c_4} = \frac{c_4(\lu-\ls)}{\kT}\,
        S_\tau \left[\lu, \avg{P\,u^\ddag(0) \, X^3(\tau)}_\alpha ; 0\right].
\end{equation}
The average over the noise can be brought inside the $S$~functional because the latter is
shorthand notation for an integral. The remaining moment can be evaluated as
\begin{align}
    \label{kappa2Moment}
    \avg{P\,u^\ddag(0) \,X^3(\tau)}_\alpha
    &= \frac{\lu}{\omb} \avg{u^\ddag(0) \,X^3(\tau)}_0 \nonumber \\
    &= 3\frac{\lu}{\omb} \avg{u^\ddag(0)\, X(\tau)}_0\,\avg{X^2(\tau)}_0.
\end{align}
The modified correlation functions that are required here are given in Eq.~\eqref{XMoments}.
Equation~\eqref{kappa2Moment} can thus be rewritten as a sum of exponentially decaying terms,
for which the $S$~functional in Eq.~\eqref{kappa2First} is easy to evaluate.
This procedure yields
\begin{equation}
    \label{kappa2Final}
    \kappa_2^{c_4} = -\frac{3c_4 \sigma^4 (\lu-\ls)^2}{4\kT \omb\lu}
        = \tfrac{3}{4} c_4 \kT \frac{\gamma^2}{\omb^3 \ls (\lu-\ls)^2}.
\end{equation}

\begin{figure}
\includegraphics{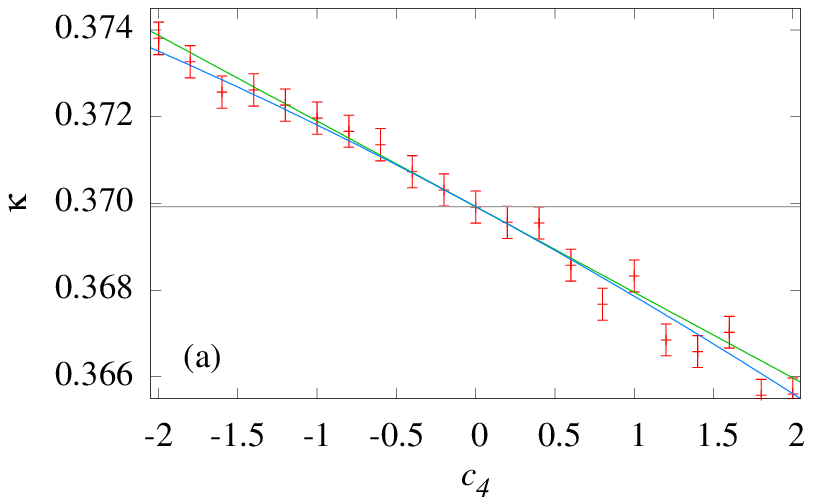}
\includegraphics{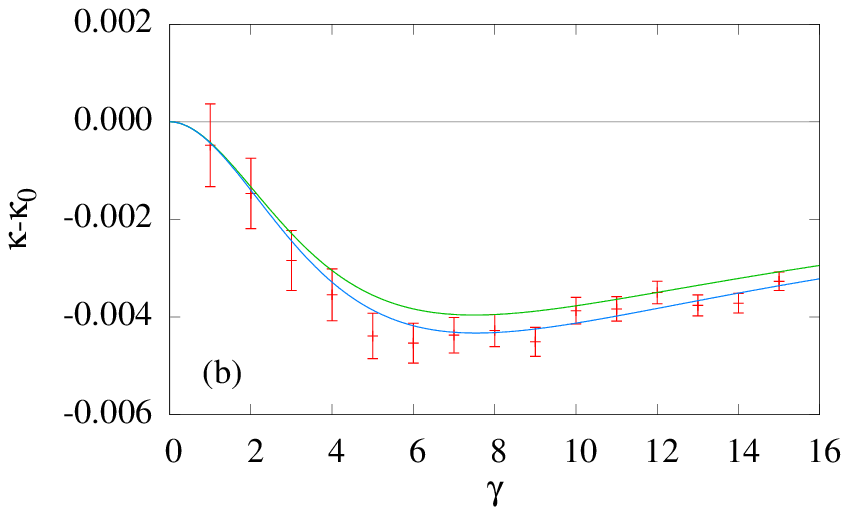}
\caption{
 Transmission factor, $\kappa$, for a one-dimensional potential with
       quartic anharmonicity, $c_4$,
       for $\omb=3$, $\kT=1$. \\
 (a)   $\kappa$ as a function of the coupling strength $c_4$ for a value
       of the damping $\gamma=7$. \\
 (b)   Difference between $\kappa$ and its Kramers approximation, $\kappa_0$,
       as a function of $\gamma$ for $c_4=2$: \\
       Numerical simulation results (red points),
       harmonic (Kramers) approximation~\eqref{kappaKramers} (gray horizontal line),
       perturbative results to first-order,
       obtained from~\eqref{kappaKramers}+\eqref{kappa2Final} (green line), and
       second-order obtained from~\eqref{kappaKramers}+\eqref{kappa2Final}+\eqref{kappa4Quartic} (blue line).}
\label{fig:kappaQuartic}
\end{figure}

This result agrees with the perturbative correction given in
Refs.~\onlinecite{Pollak93a,Talkner93,Talkner94a}.
It can be rewritten as
\begin{equation}
    \label{kappa2Quartic}
    \frac{\kappa_2^{c_4}}{\kappa_0} =
     -\frac{3}{4} \frac{c_4\, \kT}{\omb^4}
        \left(\frac{1-\mu^2}{1+\mu^2}\right)^2
\end{equation}
in terms of the dimensionless parameter $\mu=\kappa_0=\lu/\omb$ that was used in Ref.~\onlinecite{Talkner93}.
A comparison of Eq.~\eqref{kappa2Final} with  numerical results is shown in Figure~\ref{fig:kappaQuartic}.
They confirm once more that the perturbative result is correct.
The figure also shows the second-order correction in $c_4$, which can be obtained in a similar
way from Eq.~\eqref{V4quartic}. It reads
\begin{multline}
    \label{kappa4Quartic}
    \frac{\kappa_4^{c_4}}{\kappa_0}=
    -\frac{3}{32}\left(\frac{c_4\,\kT}{\omb^4}\right)^2
    \left(\frac{1-\mu^2}{1+\mu^2}\right)^4 \\
    \frac{105\mu^8+830\mu^6+1648\mu^4+770\mu^2+87}
        {(1-\mu^4)(3\mu^4+10\mu^2+3)}.
\end{multline}
In the numerical example the second-order contribution is small,
but Fig.~\ref{fig:kappaQuartic}(b) shows clearly that the second-order
perturbative result is in better agreement with the numerical data than
the first-order result.

\begin{figure}
\includegraphics{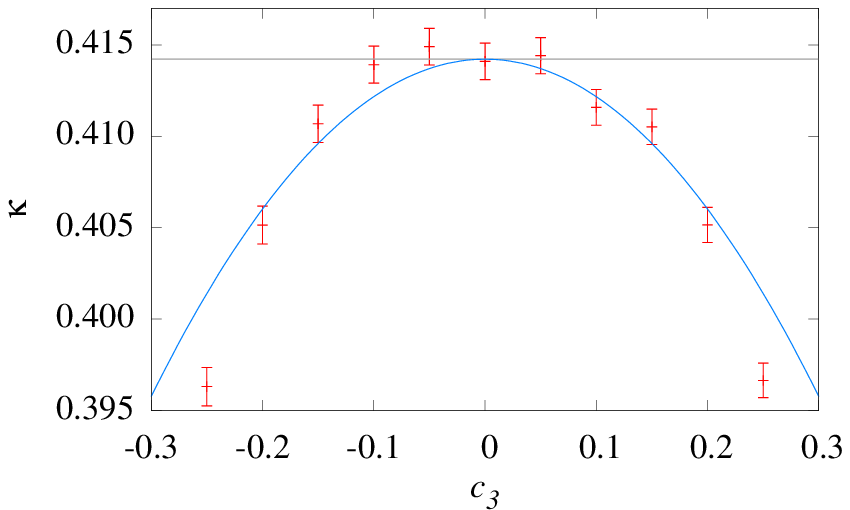}
\caption{
  Transmission factor for a one-dimensional potential with cubic
  anharmonicity, $c_3$, with $\omb=1$, $\gamma=2$, $\kT=1$: \\
  Numerical simulation results (red points),
  harmonic (Kramers) approximation~\eqref{kappaKramers} (gray horizontal line),
  perturbative results to second-order,
  obtained from~\eqref{kappaKramers}+\eqref{kappa2Cubic} (blue line).
  Notice that in this case the first-order correction is zero.}
\label{fig:kappaCubic}
\end{figure}

For a generic anharmonic potential that has a third-order term, the leading rate
correction is quadratic in~$c_3$ and can be obtained from Eq.~\eqref{kappaAvg2}
with the help of Eqns.~\eqref{V1general} and~\eqref{V2general}.
It reads
\begin{equation}
    \label{kappa2Cubic}
    \frac{\kappa_2^{c_3}}{\kappa_0}=
     -\frac{1}{6} \frac{c_3^2\, \kT}{\omb^6}
        \left(\frac{1-\mu^2}{1+\mu^2}\right)^2
        \frac{10\mu^4+41\mu^2+10}{2\mu^4+5\mu^2+2}.
\end{equation}
A comparison between Eq.~\eqref{kappa2Cubic} and numerical data is shown
in Fig.~\ref{fig:kappaCubic}. Again, the agreement is excellent.

If both cubic and quartic perturbations are present in the potential,
then the second order contribution to the Kramers' transmission factor equals
to the sum of expressions~\eqref{kappa2Quartic} and~\eqref{kappa2Cubic}.

\section{The two-dimensional case}
\label{sec:2d}

So far, our discussion of the stochastic stable and unstable manifolds
and their use has been restricted to a one-dimensional model.
Most problems of physical interest, however, have several degrees of freedom.
It is therefore crucial to show how the results obtained before
can be generalized to higher dimension.
We will carry out the generalization to two dimensions,
which requires some extensions of the previous discussion.
It will then be obvious that these techniques can equally be applied to systems
in arbitrary dimension.

We study a two-dimensional model whose dynamics is described by the
Langevin equation~\eqref{Langevin}.
We denote the configuration space coordinates as $\vec q=(x,y)$
and the corresponding velocities as $\dot{\vec q}=(v_x,v_y)$.
The friction matrix $\Gamma=\gamma I_2$ is assumed to be a scalar multiple
of the $2\times 2$ identity matrix, $I_2$.
By the fluctuation-dissipation theorem~\eqref{flucdis},
this assumption implies that the $x$ and $y$ components of the fluctuating
force are statistically uncorrelated.
For demonstration purposes we will use the anharmonic model potential
\begin{equation}
    \label{Potential2D}
    U(x,y) = -\frac 12 \omb^2 \, x^2 + \frac 12 \omega_y^2\,y^2
        + c\, x^2 y^2
\end{equation}
that has already been used in Refs.~\onlinecite{Bartsch06a,Bartsch08}.
The anharmonic perturbation in~\eqref{Potential2D} is of fourth order.
In the terminology of the previous sections, the coupling parameter~$c$ is
therefore of order $\varepsilon^2$, and rate corrections at first order in $c$
are expected.

\subsection{Invariant manifolds in higher dimension}
\label{sec:mf2d}

In a two-dimensional setting, the phase space of the
Langevin equation~\eqref{Langevin} is four-dimensional.
It can be described with coordinates $(x,y,v_x,v_y)$.
As before, the harmonic approximation of the dynamics around the
barrier can be diagonalized by introducing the coordinates
$u$ and $s$ given in Eq.~\eqref{suTransform} and
coordinates $z_1$ and $z_2$ defined by
\begin{align}
    z_1 &= \frac{v_y-\lambda_2 y}{\lambda_1-\lambda_2},  &
    z_2 &= \frac{v_y-\lambda_1 y}{\lambda_2-\lambda_1}
\end{align}
with the inverse transformation
\begin{align}
    y & = z_1+z_2,  & v_y &= \lambda_1 z_1 + \lambda_2 z_2.
\end{align}
The two additional eigenvalues
\begin{equation}
    \label{evalY}
    \lambda_\text{1,2} =
        -\frac 12 \left(\gamma \pm \sqrt{\gamma^2-4\omy^2}\right)
\end{equation}
are either real and negative or form a pair of complex conjugates with negative real parts.

The fluctuating force has two independent components $\xi_{x,\alpha}(t)$ and $\xi_{y,\alpha}(t)$,
which determine the four components of the TS trajectory
\begin{align}
    u^\ddag(t) &= \frac{1}{\lu-\ls}\,S[\lu, \xi_{x,\alpha};t], \nonumber \\
    s^\ddag(t) &= -\frac{1}{\lu-\ls}\,S[\ls, \xi_{x,\alpha};t], \nonumber \\
    z_1^\ddag(t) &= \frac{1}{\lambda_1 - \lambda_2}\,S[\lambda_1, \xi_{y,\alpha};t], \nonumber \\
    z_2^\ddag(t) &= -\frac{1}{\lambda_1-\lambda_2}\,S[\lambda_2, \xi_{y,\alpha};t]
\end{align}
that serves as a time-dependent coordinate origin. In the relative coordinates
\begin{align}
    \Delta u &= u-u^\ddag, \quad \Delta s = s-s^\ddag, \nonumber \\
    \Delta z_1 &= z_1 - z_1^\ddag, \quad \Delta z_2 = z_2 - z_2^\ddag
\end{align}
the Langevin equation is written as
\begin{align}
    \label{relEq2d}
    \Delta\dot u &= \lu \Delta u + \frac{f_x(x,y)}{\lu-\ls}, \nonumber \\
    \Delta \dot s &= \ls \Delta s - \frac{f_x(x,y)}{\lu-\ls}, \nonumber \\
    \Delta \dot z_1 &= \lambda_1 \Delta z_1 + \frac{f_y(x,y)}{\lambda_1-\lambda_2}, \nonumber \\
    \Delta \dot z_2 &= \lambda_2 \Delta z_2 - \frac{f_y(x,y)}{\lambda_1-\lambda_2},
\end{align}
where $f_x$ and $f_y$ denote the anharmonic parts of the mean force:
\begin{align*}
    -\frac{\partial U}{\partial x} &= \omb^2 x + f_x(x,y), \\
    -\frac{\partial U}{\partial y} &= -\omy^2 y + f_y(x,y).
\end{align*}
The differential equations~\eqref{relEq2d} are coupled by the conditions
\begin{align*}
    x &= x^\ddag + \Delta u + \Delta s, \\
    y &= y^\ddag + \Delta z_1 + \Delta z_2.
\end{align*}

As in the one-dimensional case, the equations of motion~\eqref{relEq2d}
decouple and become time-independent in the harmonic limit, $f_x=f_y=0$,
and the relevant phase space structures can easily be described in this case.
Among the eigenvalues in Eq.~\eqref{relEq2d}, $\lu$ is positive,
while the other three have negative real parts.
Consequently, the TS~trajectory has a one-dimensional unstable
manifold and a three-dimensional stable manifold.
The stable manifold separates reactive from non-reactive regions of phase space.
The dimension of the unstable manifold, by contrast, is too low to separate
distinct regions in the four-dimensional phase space.
The invariant manifolds cannot therefore be used to distinguish trajectories
with different behaviors in the remote past, but the stable manifold
can be used to predict the fate of a trajectory in the future.
Thus, in arbitrary dimension the invariant manifolds provide precisely the
diagnostic capabilities that are needed for rate calculations.

We are particularly interested in trajectories that start on the DS $x=0$.
This is a three-dimensional surface with coordinates $(v_x,y,v_y)$,
embedded in the four-dimensional phase space.
It intersects the three-dimensional stable manifold in a two-dimensional surface
that separates reactive from non-reactive trajectories within the DS.
We will call that two-dimensional surface the separatrix,
and it depends on the realization of the noise.

\begin{figure}
\includegraphics[width=245pt]{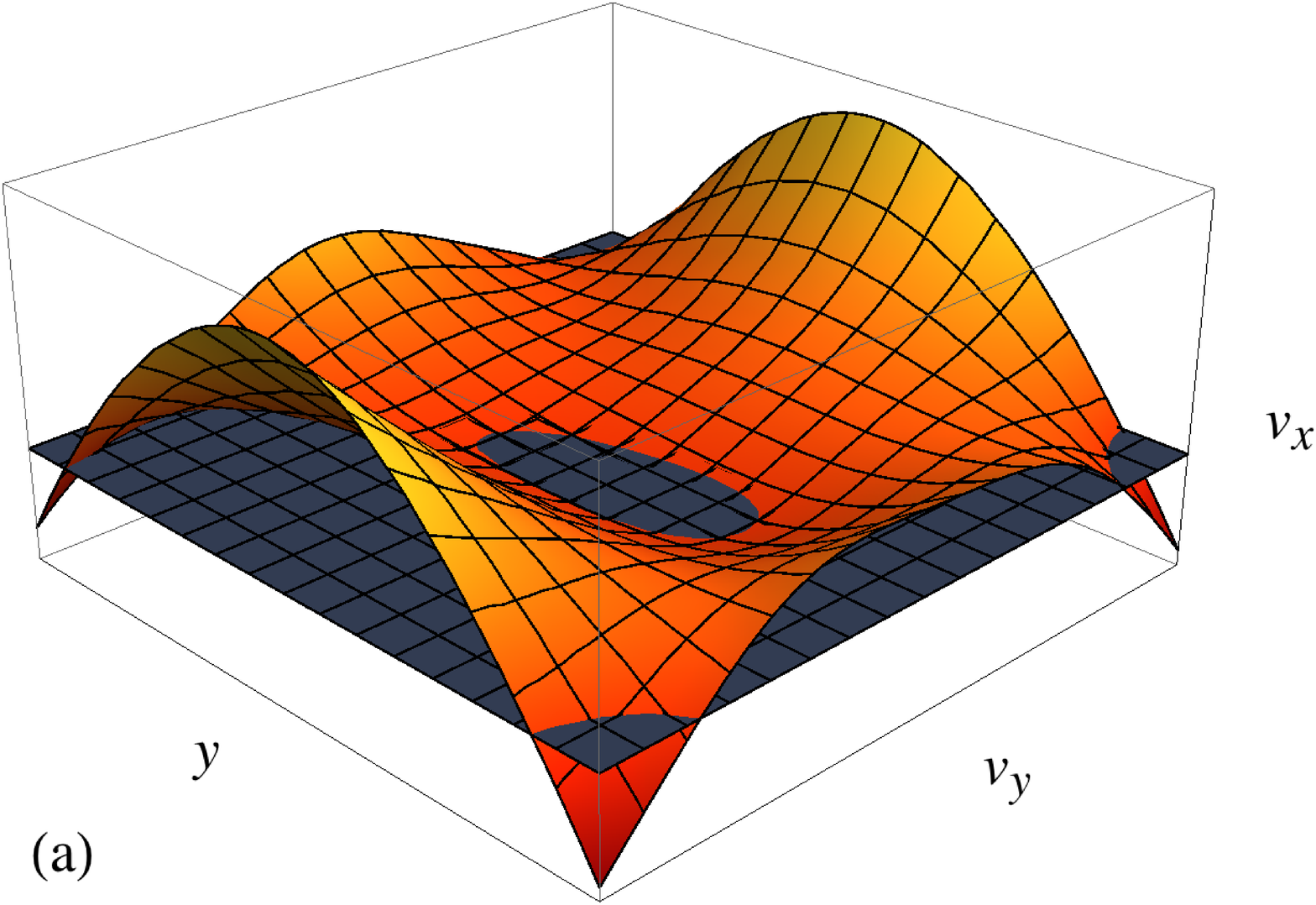}
\includegraphics[width=245pt]{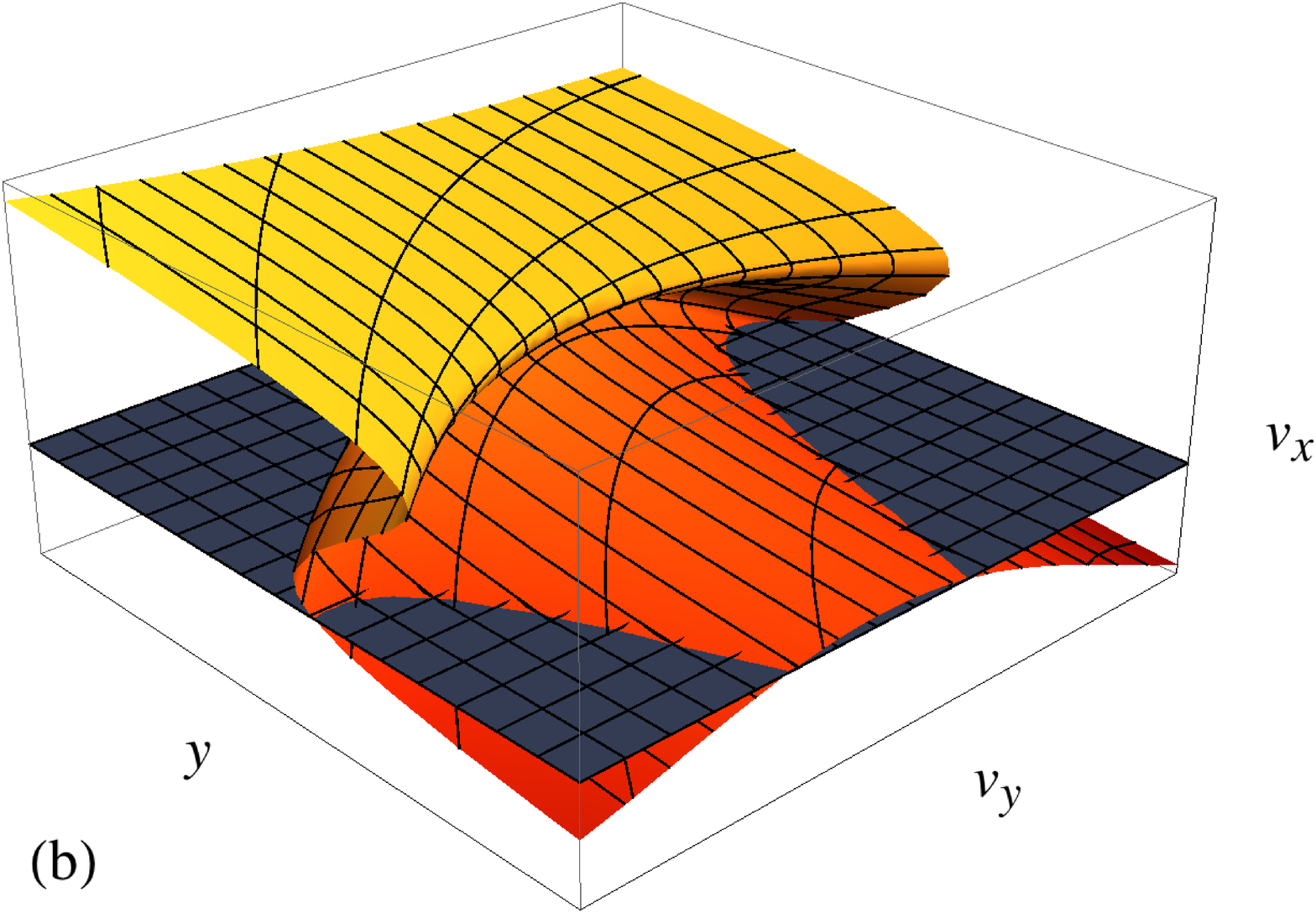}
\caption{
 Schematic representation of the separatrix within the dividing surface $x=0$.
 For a harmonic barrier the separatrix is a plane (gray in both panels).
 (a) For a weakly anharmonic barrier the separatrix can be parameterized by a function
     $V^\ddag(y,v_y)$.
     Trajectories with initial condition $v_x > V^\ddag(y,v_y)$ are reactive.
 (b) If anharmonicities are strong, the separatix cannot be described by a single critical
     velocity, $V^\ddag$.
}
\label{fig:separatrix}
\end{figure}

On physical grounds, we expect a trajectory to be reactive if its initial velocity $v_x$
is sufficiently high.
The critical velocity $V^\ddag$ that separates reactive from non-reactive trajectories
depends, in general, on the transverse coordinates $y$ and $v_y$.
In the harmonic limit, the critical velocity is given by~\eqref{V0} and is independent
of these transverse coordinates.
The separatrix $v_x=V^\ddag$ is therefore a plane within the DS that is parallel
to the $y$-$v_y$ plane.
When anharmonicities are taken into account, the separatrix is deformed from this plane in a
stochastically time-dependent way, as indicated schematically in Figure~\ref{fig:separatrix}(a).
Nevertheless, we will still be able to describe the separatrix by specifying a critical velocity
that depends on the transverse coordinates.
In Section~\ref{sec:pert2d} a perturbative expansion for the function~$V^\ddag(y,v_y)$
will be developed.

\begin{figure}
 \includegraphics{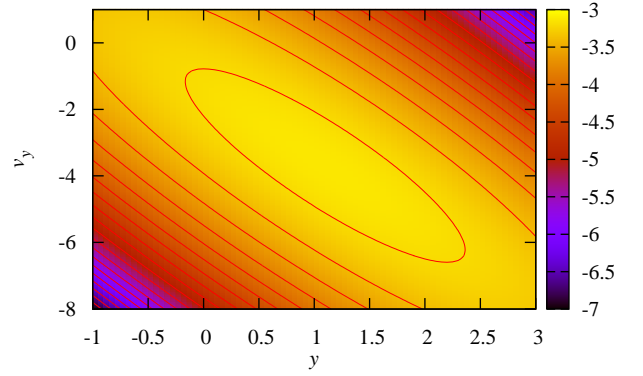}
 \caption{Critical velocity as a function of the transverse coordinates
   for one realization of the noise for the two-dimensional model potential~\eqref{Potential2D}
   for $\omega_x=1$, $\omega_y=1.5$, $\gamma=2$, $c=0.2$, $\kT=1$.
   Contour spacing is 0.2 and the central contour value is -3.2.}
 \label{fig:VCrit2d}
\end{figure}

It is instructive to study the actual shape of the separatrix in a representative example.
Figure~\ref{fig:VCrit2d} shows the critical velocity as a function of transverse coordinates
for one realization of the noise for the two-dimensional model potential~\eqref{Potential2D}.
The critical velocity takes a maximum that is noticeably displaced from the origin $y=v_y=0$.
At the maximum, the critical velocity is closest to its harmonic value,
which in this case is approximately $-3.01$.
For all values of the transverse coordinates, the critical velocity is below the
harmonic approximation value.
Moreover, it decays steeply away from the maximum, so that deviations from the harmonic
approximation are large for most values of the coordinates.
As the critical velocity appears in the exponent in the rate formula~\eqref{kappaVel}
--- which will be generalized to higher dimension in Eq.~\eqref{kappaVel2d} ---,
it is expected that anharmonic effects on the critical velocity leads to large rate corrections.

If the barrier is strongly anharmonic it cannot be guaranteed, in general, that the
separatrix can be parameterized by the transverse coordinates $y$ and $v_y$.
In a situation as that indicated in Fig.~\ref{fig:separatrix}(b), the separatrix
is described by a multivalued function of the transverse coordinates.
It cannot be characterized by a single critical velocity.
As expected.trajectories at low $v_x$ are nonreactive,
and those at somewhat larger $v_x$ are reactive. However, at certain values of $y$ and $v_y$,
there is an interval
at yet higher $v_x$ that also contains nonreactive trajectories.
A scenario like this obviously requires very strong anharmonic effects,
and this is only be achieved for large values of the transverse coordinates.
But at these conditions, it is doubtful whether a TST-like treatment with a single
rate-determining saddle point is appropriate at all.
We will therefore neglect this possibility and assume the existence of a single
critical velocity.

\subsection{Determination of the stable manifold}
\label{sec:pert2d}

As a basis for the perturbative expansion, we formally solve the differential
equations~\eqref{relEq2d} in terms of $S$~functionals by
\begin{align}
    \label{intEq2d}
    \Delta u(t) &= \frac{1}{\lu-\ls}\,S[\lu,f_x(x,y); t], \nonumber \\
    \Delta s(t) &= \Delta s(0) e^{\ls t} - \frac{1}{\lu-\ls}\,\bar S[\ls,f_x(x,y); t], \nonumber \\
    \Delta z_1(t) &= \Delta z_1(0) e^{\lambda_1 t} + \frac{1}{\lambda_1-\lambda_2}\,
                \bar S[\lambda_1, f_y(x,y); t], \nonumber \\
    \Delta z_2(t) &= \Delta z_2(0) e^{\lambda_2 t} - \frac{1}{\lambda_1-\lambda_2}\,
                \bar S[\lambda_2, f_y(x,y); t].
\end{align}
These integral equations are entirely analogous to Eqns.~\eqref{uEqInt} and~\eqref{sEqInt},
and they are coupled by
\begin{align*}
    x &= x^\ddag + \Delta u + \Delta s, \\
    y &= y^\ddag + \Delta z_1 + \Delta z_2.
\end{align*}
A trajectory satisfying~\eqref{intEq2d} automatically lies on the stable manifold.
To find the critical velocity, Eqns.~\eqref{intEq2d} needs to be solved under the
condition that the trajectory starts in the DS $x=0$
and at the prescribed transverse coordinates $y(0)$ and $v_y(0)$.

We will solve Eqns.~\eqref{intEq2d} by an iterative procedure as in~\eqref{perturbRecur}.
As before, the initial condition $\Delta s(0)$ must be adapted in every step in order
to enforce the condition $x(0)=0$.
By contrast, the transverse initial conditions $\Delta z_1(0)$ and $\Delta z_2(0)$
are fixed once and for all by imposing the condition that
\begin{align*}
    y(0) &= y^\ddag(0) + \Delta z_1(0) + \Delta z_2(0), \\
    v_y(0) &= v_y^\ddag(0) + \lambda_1 \Delta z_1(0) + \lambda_2 \Delta z_2(0)
\end{align*}
take the desired values.
The critical velocity is finally obtained from Eq.~\eqref{VCritU}.

Our perturbation expansion is centered around the harmonic approximation to a
trajectory on the stable manifold, given by Eq.~\eqref{XDef}
\[
    X(t) = x^\ddag(t) - x^\ddag(0) e^{\ls t}
\]
and
\begin{align}
    \label{YDef}
    Y(t) &= y^\ddag(t) + \Delta z_1(0) e^{\lambda_1 t}
        + \Delta z_2(0) e^{\lambda_2 t}.
\end{align}
The latter can be split according to
\begin{equation}
    \label{YSplit}
    Y(t) = Y_\alpha(t) + Y_\bot(t)
\end{equation}
into one part
\[
    Y_\alpha(t) = y^\ddag(t) - z_1^\ddag(0) e^{\lambda_1 t} - z_2^\ddag(0) e^{\lambda_2 t}
\]
that depends on the realization of the noise but not on the initial conditions,
and another
\[
    Y_\bot(t) = z_1(0) e^{\lambda_1 t} + z_2(0) e^{\lambda_2 t}
\]
that depends on the initial conditions but not on the noise.

We will now apply the general theory to the model potential~\eqref{Potential2D}.
Our aim is to expand the coordinates
\begin{align*}
    x(t) &= X(t) + c \,\Delta x_1(t) + c^2 \Delta x_2(t) + \dots, \\
    y(t) &= Y(t) + c \,\Delta y_1(t) + c^2 \Delta y_2(t) + \dots
\end{align*}
in powers of the anharmonicity parameter~$c$.
For expansions of other quantities, such as
\[
    V^\ddag = V^\ddag_0 + c V^\ddag_1 + c^2 V^\ddag_2 + \dots,
\]
a similar notation will be used. The anharmonic forces are given by
\begin{align*}
    f_x &= -2c\,xy^2 \\
        &= -2c\,XY^2 - 2c^2(Y^2\Delta x_1 + 2 XY\,\Delta y_1) + \dots,\\
    f_y &= -2c\,x^2y \\
        &= -2c\,X^2Y - 2c^2(2XY\,\Delta x_1 + X^2 \Delta y_1) + \dots.
\end{align*}

\begin{figure}
 \includegraphics{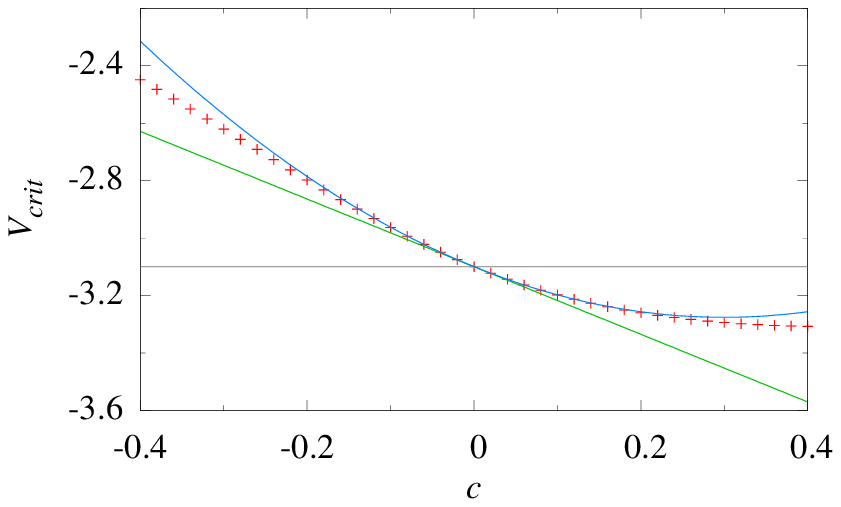}
 \caption{Critical velocity for one realization of the noise for the two-dimensional
   model potential ~\eqref{Potential2D} with $\omega_x=1$, $\omega_y=1.5$, $\gamma=2$, $\kT=1$,
   for an initial condition $y=0$, $v_y=0$. \\
   Numerical simulation results (red crosses), harmonic approximation~\eqref{V0} (gray horizontal line),
   perturbative results to first-order~\eqref{V0}+\eqref{VCrit2d1} (green straight line) and
   second-order~\eqref{V0}+\eqref{VCrit2d1}+\eqref{VCrit2d2} (blue line).}
\label{fig:VCrit2D_VarC}
\end{figure}

\begin{figure}
 \includegraphics{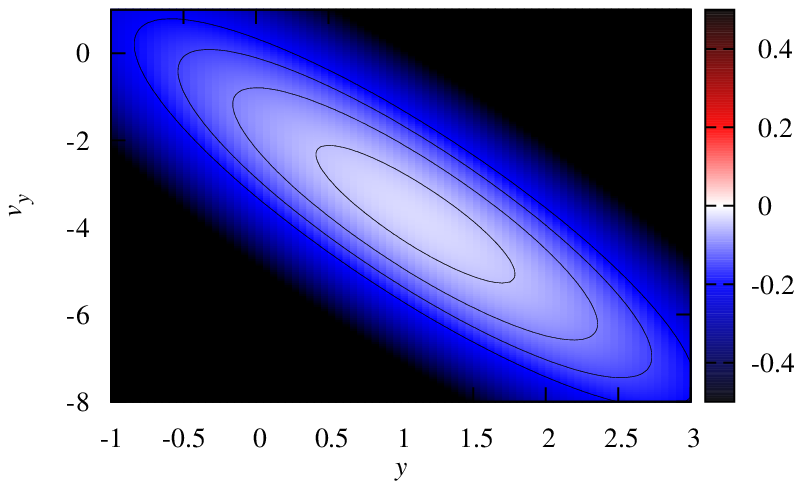}
 \includegraphics{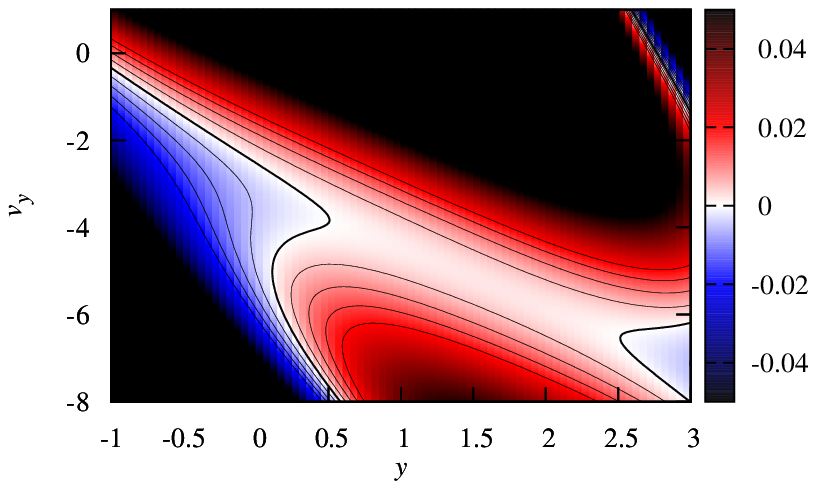}
 \includegraphics{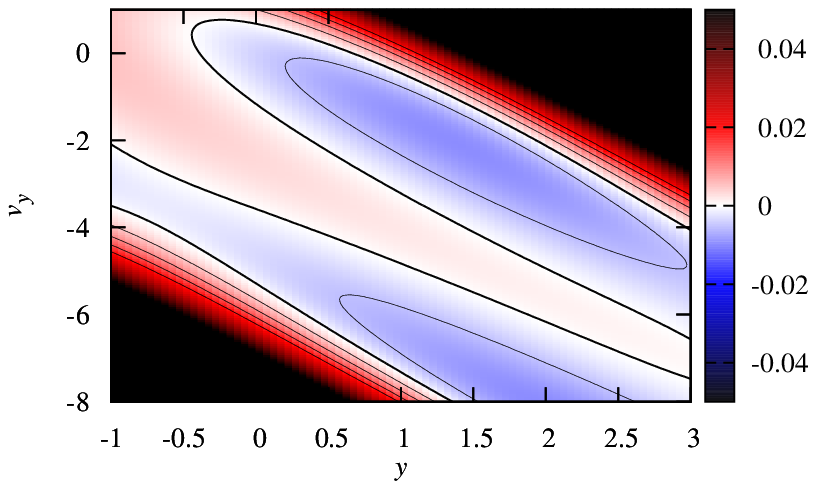}
 \caption{Difference between numerically calculated critical velocity and perturbative expansions.
   Noise sequence and parameter values as in Fig.~\ref{fig:VCrit2d}.
   (a) Harmonic approximation.
   (b) First-order perturbation theory.
   (c) Second order perturbation theory.
   Contour spacing is 0.05 in (a), 0.005 in (b) and (c).
   Note that the color scale is also stretched by a factor~10 in (a).
}
\label{fig:VCrit2dPert}
\end{figure}

In the first step of the iteration we find
\begin{align}
  \label{eq70}
    \Delta u_1(t) &= \frac{1}{\lu-\ls}S[\lu,f_{x,1};t] \nonumber \\
        &= -\frac{2}{\lu-\ls} S[\lu,XY^2;t],
\end{align}
where $f_{x,n}$ is the coefficient of $f(x)$ of order $c^n$.
From Eq.~\eqref{eq70} we get
\begin{align}
    \label{VCrit2d1}
    V^\ddag_1 &= (\lu-\ls)\Delta u_1(0) \nonumber \\
        &= -2 S[\lu, XY^2; 0].
\end{align}
The remaining coordinates need only be calculated if the second-order correction
for the critical velocity is desired.
We then obtain
\begin{align*}
    \Delta s_1(t) &= -\Delta u_1(0) e^{\ls t} + \frac{2}{\lu-\ls} \bar S[\ls, XY^2;t], \\
    \Delta z_1(t) &= -\frac{2}{\lambda_1-\lambda_2} \bar S[\lambda_1, X^2Y; t], \\
    \Delta z_2(t) &= +\frac{2}{\lambda_1-\lambda_2} \bar S[\lambda_2, X^2Y; t].
\end{align*}
Finally, with the aid of
\begin{align*}
    \Delta x_1 &= \Delta u_1 + \Delta s_1, \\
    \Delta y_1 &= \Delta z_1 + \Delta z_2,
\end{align*}
we can calculate
\[
    \Delta u_2(t) = \frac{1}{\lu-\ls}S[\lu,f_{x,2};t].
\]
The resulting expression  reduces to
\begin{widetext}
\begin{align}
    \label{VCrit2d2}
    V^\ddag_2 &= -4 \, S_\tau\Biggl[\lu, \quad
        \frac{Y^2(\tau)}{\lu-\ls} \biggl(
            S\left[\lu,XY^2;0\right] e^{\ls\tau}
            - S\left[\lu, XY^2;\tau\right]
            + \bar S\left[\ls, XY^2;\tau\right] \biggr)
            \nonumber \\
        & \hspace{15em} + 2\,\frac{X(\tau)\,Y(\tau)}{\lambda_1-\lambda_2} \biggl(
            \bar S\left[\lambda_2, X^2 Y;\tau\right]
            - \bar S\left[\lambda_1, X^2Y;\tau\right]
        \biggr)
        ; \;\;0 \Biggr].
\end{align}
\end{widetext}
Figure~\ref{fig:VCrit2D_VarC} shows the value of the critical velocity for one realization 
of the noise for the two-dimensional model potential ~\eqref{Potential2D} as a function of 
the coupling strengh, $c$, for the initial condition $y=0$, $v_y=0$. 
It is compared to perturbative results up 
to second order. 
As it can be seen, our perturbative results agree very well with those obtained numerically, 
thus showing the efficiency of our method. 
To further analyze the performance of our method,
we show in Fig.~\ref{fig:VCrit2dPert} the difference between the numerically calculated 
critical velocity and the value obtained with our perturbative expansions for different
values of the transverse coordinates, where it is clearly seen that it sensibly reduces 
as the order of the perturbation is increased.

\subsection{Reaction rate expressions}
\label{sec:rate2d}

The simple expression~\eqref{kappaVel} for the transmission coefficient in terms
of the critical velocity can easily be generalized to higher dimension.
To achieve this, we start again from Eq.~\eqref{kappaFlux}.
Note first that in the denominator of Eq.~\eqref{kappaFlux} the average
over the transverse coordinates has no effect since the TST approximation
to the characteristic function does not depend on them.
In the numerator, we use again the form~\eqref{chiVcrit1D} of the
characteristic function and carry out the average over $v_x$ as before, to obtain
\begin{equation}
    \label{kappaVel2d}
    \kappa = \avg{ \exp\left(-\frac{V^{\ddag2}}{2\kT}\right) }_{\alpha\bot}.
\end{equation}
In this expression the average over the transverse coordinates,
which is indicated by subscript $\bot$,
cannot be carried out immediately because the critical velocity
depends on the transverse coordinates.

Equation~\eqref{kappaVel2d} represents the simplest conceivable generalization
of Eq.~\eqref{kappaVel}.
It is remarkable that no modifications, beyond the additional average
over the transverse coordinates, are required.
This is only possible because no anharmonic corrections are required for the
denominator in Eq.~\eqref{kappaFlux}.

In the case of the model potential~\eqref{Potential2D}, the distribution~\eqref{pPerp}
of the transverse coordinates is given by
\begin{equation}
    \label{pPerp2d}
    p_\bot(y,v_y) = \frac{1}{Z}\, \exp\left(-\frac{v_y^2+\omy^2 y^2}{2\kT} \right),
\end{equation}
i.e., it is a Gaussian distribution.
The functions $X$ and $Y$ will then both have a Gaussian distribution,
which allows us to evaluate the rate corrections by the method of Sec.~\ref{sec:moments}.
For any expression involving $u^\ddag(0)$, $X$ and $Y$, we write
\begin{equation}
    \label{modAvg2d}
    \avg{P(\dots)}_{\alpha\bot} = \frac{\lu}{\omb}\avg{\dots}_{0\bot}
\end{equation}
as in Eq.~\eqref{avgMod}.
The average over the initial conditions is not involved in the transition from the
noise average to the distorted average with correlation function~\eqref{modSigmaComp},
because the noise and the initial conditions are uncorrelated.

Once we have a perturbative expansion of the critical velocity of the form~\eqref{Vexpand},
expressions~\eqref{kappaAvg} can be used for the expansion of the transmission factor.
The only required modification being to replace noise averages by averages over
noise and the transverse coordinates.

Assuming a general anharmonic potential of the form
\[
    U(0,y) = \frac{1}{2} \omy^2 y^2 + U_\text{anh}(y),
\]
where $U_\text{anh}(y)$ contains terms at least of third order in $y$,
i.e. at least of first order in the expansion parameter~$\varepsilon$,
it can be treated perturbatively in the current framework.
The distribution function of the transverse coordinates can then be expanded as
\begin{multline}
    \label{pPerpExp}
    p_\bot(y,v_y) = \frac{1}{Z}\, \exp\left(-\frac{v_y^2+\omy^2 y^2}{2\kT} \right) \\
        \times\left(1+\varepsilon\,a_1(y) + \varepsilon^2\,a_2(y) + \dots \right)
\end{multline}
with suitable coefficients~$a_i$ that are polynomials in~$y$ of degree at most~$i$.
We assume that the partition function~$Z$ in Eq.~\eqref{pPerpExp} is the same
as in the Gaussian distribution~\eqref{pPerp2d}, and any corrections
to the partition function that arise from the anharmonicity of the potential
have been included in the expansion coefficients~$a_i(y)$.

Using symbol $\Perp$ to denote an average over the Gaussian distribution~\eqref{pPerp2d}
of initial conditions, we can write
\begin{align*}
    \kappa &= \avg{\exp\left(-\frac{V^{\ddag2}}{2\kT}\right)}_{\alpha\bot} \\
        &= \avg{\exp\left(-\frac{V^{\ddag2}}{2\kT}\right)\times\left(1+\varepsilon\,a_1(y)
        + \varepsilon^2\,a_2(y) + \dots \right)}_{\alpha\Perp} .
\end{align*}
The expansion~\eqref{Vexpand} of the critical velocity then allows us to expand the
exponential, thus obtaining
\[
    \kappa = \kappa_0 + \varepsilon\kappa_1 + \varepsilon^2\kappa_2 + \dots
\]
with
\begin{subequations}
\label{kappaAvg2d}
\begin{align}
    \kappa_0 &= \avg{ P }_{\alpha\Perp}, \label{kappaAvg2d0} \\
    \kappa_1 &= -\frac{1}{\kT} \avg{ P V_0^\ddag V_1^\ddag}_{\alpha\Perp} +
        \avg{P\,a_1(y)}_{\alpha\Perp}, \label{kappaAvg2d1}\\
    \kappa_2 &= \frac{1}{2(\kT)^2} \avg{P V_0^{\ddag2} V_1^{\ddag2}}_{\alpha\Perp}
        - \frac{1}{\kT} \avg{P V_0^\ddag V_2^\ddag}_{\alpha\Perp} \nonumber \\
        & \quad - \frac{1}{2\kT} \avg{P V_1^{\ddag2}}_{\alpha\Perp}
            -\frac{1}{\kT}\avg{P\,V^\ddag_0V^\ddag_1 a_1(y)}_{\alpha\Perp}
            + \avg{P\,a_2(y)}_{\alpha\Perp} \label{kappaAvg2d2}
\end{align}
\end{subequations}
wehre again the abbreviation~\eqref{EDef} has been used.
The remaining averages are Gaussian averages that can be evaluated, as before,
by first converting the noise average into a distorted Gaussian average
via~\eqref{modAvg2d}, and then using Isserlis' theorem.

Because the factor~$P$ is independent of the initial conditions,
we obtain from~\eqref{kappaAvg2d0}
\begin{align*}
    \kappa_0 = \avg{P}_{\alpha} = \frac{\lu}{\omb},
\end{align*}
the Kramers result.
Similarly, the expressions $\avg{P\,a_i(y)}_{\alpha\Perp}$,
that occur in all correction terms, can be simplified to
\[
    \avg{P\,a_i(y)}_{\alpha\Perp} = \avg{P}_\alpha\,\avg{a_i(y)}_\Perp
        = \frac{\lu}{\omb}\,\avg{a_i(y)}_\Perp.
\]

\subsection{Correlation functions}
\label{sec:moments2d}

To evaluate corrections to the transmission factor in Eq.~\eqref{kappaAvg2d}
using Isserlis' theorem, the correlation functions $\avg{w_1 w_2}_{0\Perp}$,
where $w_1$ and $w_2$ are one of $u^\ddag(0)$, $X(t)$, $Y(t)$, and $y(0)$,
are needed.
(The initial condition $y(0)$ was written without its time argument in Sec.~\ref{sec:rate2d}.
For the sake of clarity we will now include it again.)

Because the $x$ and $y$ components of the fluctuating force are uncorrelated,
all correlation functions involving one of either $u^\ddag(0)$ or $X(t)$
and one of either $Y(t)$ or $y(0)$ must vanish.
Furthermore, since $u^\ddag(0)$ and $X(t)$ do not depend on initial conditions,
\[
    \avg{u^\ddag(0) X(t)}_{0\Perp} = \avg{u^\ddag(0) X(t)}_0
\]
and
\[
    \avg{X(t)X(t')}_{0\Perp} = \avg{X(t)X(t')}_0
\]
are given by Eq.~\eqref{XMoments}.

Concerning the initial conditions, it can be read off from the distribution
function~\eqref{pPerp2d} that
\begin{equation}
    \label{yAvg}
    \avg{y(0)^2}_{0\Perp} = \frac{\kT}{\omy^2}.
\end{equation}
(The average over the distorted noise distribution does not have any effect.)
We can also see that
\begin{equation}
    \label{vyAvg}
    \avg{v_y(0)^2}_{0\Perp} = \kT
    \qquad \text{and} \qquad
    \avg{y(0)\,v_y(0)}_{0\Perp} = 0.
\end{equation}
These results further yield
\begin{align}
    \label{yYavg}
    \avg{y(0)\,Y(t)}_{0\Perp} &=
        \avg{y(0)\,Y_\perp(t)}_{0\Perp} \nonumber \\
        &= \avg{y(0)z_1(0)}_{0\Perp} e^{\lambda_1 t} +
            \avg{y(0)z_2(0)}_{0\Perp} e^{\lambda_2 t} \nonumber  \\
        &= \frac{\kT}{\omy^2(\lambda_1-\lambda_2)}
            \left(\lambda_1 e^{\lambda_2 t} - \lambda_2 e^{\lambda_1 t}\right).
\end{align}

Finally, the  autocorrelation function of $Y(t)$ can be decomposed,
with the help of the split~\eqref{YSplit}, into
\begin{equation}
    \label{YAvgSplit}
    \avg{Y(t)Y(t')}_{0\Perp}
        = \avg{Y_\alpha(t) Y_\alpha(t')}_\alpha + \avg{Y_\perp(t) Y_\perp(t')}_\Perp
\end{equation}
because
\[
    \avg{Y_\alpha(t) Y_\perp(t')}_{0\Perp} = \avg{Y_\alpha(t)}_0\,\avg{Y_\perp(t')}_\perp = 0
\]
and
\[
    \avg{Y_\alpha(t) Y_\alpha(t')}_0 = \avg{Y_\alpha(t) Y_\alpha(t')}_\alpha.
\]
To evaluate the first term in Eq.~\eqref{YAvgSplit}, the correlation function of the
components $z_i^\ddag(t)$ of the TS trajectory, given in Ref.~\onlinecite{Bartsch05c},
are needed.
The second term can be evaluated with the help of Eqns.~\eqref{yAvg} and~\eqref{vyAvg}.
Finally, one arrives to the following simple result
\begin{equation}
    \label{YAvg}
    \avg{Y(t)Y(t')}_{0\Perp}  = \frac{\kT}{\omy^2-\lambda_1^2}\, e^{\lambda_1 |t-t'|}
     +  \frac{\kT}{\omy^2-\lambda_2^2}\, e^{\lambda_2 |t-t'|}.
\end{equation}
With that we have found all correlation functions that we will need to calculate the
rate corrections.

\subsection{Rate corrections}
\label{sec:rate2dresults}

Let us now derive an expansion of the transmission factor for
the case of the anharmonic model potential~\eqref{Potential2D},
\begin{equation}
    \kappa = \kappa_0 +c\, \kappa_1 + c^2 \kappa_2 + \dots,
\end{equation}
in powers of the coupling parameter~$c$.
As discussed earlier, this corresponds to an expansion in powers of $\varepsilon^2$,
and the rate formulas~\eqref{kappaAvg2d} with $a_i(y)=0$ can be used.

The first correction term is
\begin{align}
    \kappa_1 &= -\frac{1}{\kT}\,\frac{\lu}{\omb}
                 \avg{V^\ddag_0 V^\ddag_1}_{0\Perp} \nonumber \\
        &= -\frac{2}{\kT}\,\frac{\lu}{\omb}\,(\lu-\ls)\, S_\tau\left[ \lu,
                \avg{u^\ddag(0) X(\tau) Y^2(\tau)}_{0\Perp}     ; 0 \right].
\end{align}
The remaining average can be simplified to
\[
    \avg{u^\ddag(0) X(\tau) Y^2(\tau)}_{0\Perp}
    = \avg{u^\ddag(0) X(\tau)}_0 \avg{Y^2(\tau)}_{0\Perp}.
\]
The results of Sec.~\ref{sec:moments2d} give a sum of exponentially decaying terms
for this expression,
so that the $S$~functional can be evaluated as in the one-dimensional case.
In terms of the dimensionless parameters $\mu=\kappa_0=\lu/\omb$, that was already
used above, and $\nu=\omy/\omb$ the rate correction reads
\begin{equation}
    \label{kappa2d_1}
    \kappa_1 = -\frac{\gamma\,\kT}{\omb^5}\,\frac{\mu^2}{(1+\mu)^2 \nu^2}.
\end{equation}
The second-order correction can be obtained in a similar way.
After tedious calculations, one finally arrives at
\begin{widetext}
\begin{align}
    \label{kappa2d_2}
    \kappa_2 &= \frac{\mu\,(\kT)^2}{6 \omb^8}\Bigg(
        \frac{96 (\mu ^2-1)^2}{(\mu ^2+1)^2 (\mu^2-4 \nu ^2-2)}
        -\frac{6}{(\mu ^2+1) (\nu^2+1)}
            -\frac{16}{(2 \mu ^2+1) (3 \mu ^2+4 \nu^2+6)}
        +\frac{9 (\mu ^2-1) (3 \mu ^4+8 \mu^2+1)}{(\mu ^2+1)^3 \nu ^4} \nonumber \\
        & \qquad\qquad\qquad
        +\frac{64 \mu^8}{(\mu ^2+1)^2 (\mu ^2+2) (\mu ^4-2 \mu^2 (2 \nu ^2+1)-8)}
        -\frac{96 (\mu ^4+2 \mu^2-1) \mu ^6}{(\mu ^2+1)^4 (\mu ^4-2 \mu ^2(2 \nu ^2+1)-3)} \nonumber \\
        & \qquad\qquad\qquad
        +\frac{192 \mu ^6}{(2 \mu ^6+7\mu ^4+7 \mu ^2+2) (\mu ^2 (4 \nu^2+6)+3)}
        +\frac{2 (16 \mu ^{12}-24 \mu ^{10}-139\mu ^8-75 \mu ^6+77 \mu ^4+111 \mu^2+34)}
              {(\mu^2+1)^4 (2 \mu ^4+5 \mu ^2+2) \nu ^2}
   \Bigg).
\end{align}
\end{widetext}

\begin{figure}
 \includegraphics{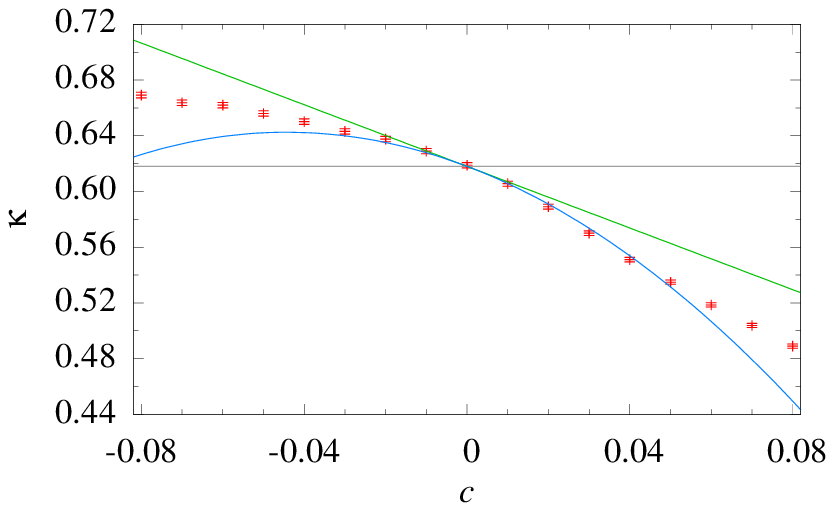}
 \caption{Transmission factor for the two-dimensional model potential~\eqref{Potential2D}
   as a function of coupling strength, $c$, for $\omb=1$, $\omy=0.5$, $\kT=1$, $\gamma=1$: \\
   Numerical simulation results (red points),
   harmonic (Kramers) approximation~\eqref{kappaKramers} (gray horizontal line),
   perturbative results to first-order, obtained from~\eqref{kappaKramers}+\eqref{kappa2d_1} (green line), and
   second-order obtained from~\eqref{kappaKramers}+\eqref{kappa2d_1}+\eqref{kappa2d_2} (blue line).}
\label{fig:kappa2d}
\end{figure}

A numerical example is shown in Fig.~\ref{fig:kappa2d}.
The second-order corrections to the transmission coefficient are small, so that a large number
of trajectories needs to be included in the numerical calculation of the rate.
Nevertheless, it is clear that the perturbative expressions~\eqref{kappa2d_1}
and~\eqref{kappa2d_2} describe the rate correctly.

\section{Concluding Remarks}
\label{sec:conc}

TST and related schemes have been widely used for rate calculations for a long time.
For reactions that occur in solution, recrossings of the DS pose a major difficulty
in such calculations.
Many approaches try to overcome this problem by choosing the DS judiciously.
By contrast, the method developed here is insensitive to the choice of this surface.
The simplest choice of DS,
which was taken here,  also leads to the simplest calculation of the critical velocity.
The use of a
different DS would require a redefinition of the critical velocity to describe
its intersection with the stable manifold, but this can be achieved with only minor modifications
to the iteration procedure for the critical velocity.
After that, any DS that lies within the barrier region would give the same rate.

This independence of the DS is achieved by two crucial features of our method.
First, the dynamics are described in phase space, rather than in configuration space,
and modern geometric methods are used in our study.
Second, we focus on invariant geometric structures that are determined by the dynamics,
rather than in structures, such as the DS, that are arbitrarily imposed
by the researcher.
The present results indicate that similar results apply to reactive systems that are coupled
to their environments, i.e., TST should focus on invariant structures in phase space.

The focus of the present paper has been on analytic perturbation theory for the rate
corrections on an anharmonic barrier.
The different steps of this calculation have different levels of complexity.
The critical velocity, which encodes the location of the invariant manifold,
is very simple to calculate with the iteration scheme described here.
Moreover, it can easily be extended to higher orders.
By contrast, the evaluation of the averages that yield the rate corrections is laborious.
While straightforward in principle, it requires the calculation of a large number of
exponential integrals something that, even for some of the results presented here,
is only feasible with the help of a computer algebra system,
\emph{Mathematica}\cite{Mathematica} in our case.

The crucial step that sets the current method apart from earlier algorithms
is the calculation of the stable manifold and the critical velocity.
It is encouraging, therefore, that this most important step of the calculation is
also the easiest.
This observation further suggests that to obtain an efficient algorithm to compute rates,
the calculation of the stable manifold should be combined with numerical methods
for the computation of averages.
We will report on such combinations in a forthcoming publication.

\section*{Acknowledgements}
This work has been supported by the MCINN (Spain) under projects
MTM2009-14621 and CONSOLIDER 2006--32 (i--Math).
FR gratefully acknowledges a doctoral fellowship the UPM and the hospitality
of the members of the School of Mathematics at Loughborugh University,
where part of this work was done.


%

\end{document}